\pdfoutput=1 
\documentclass[%
reprint,
showpacs,preprintnumbers,
 amsmath,amssymb,
 aps,
prb,
floatfix,
]{revtex4-1}

\usepackage{graphicx}
\usepackage{dcolumn}
\usepackage{bm}
\usepackage{hyperref}
\usepackage{multirow}
\usepackage{amsmath,accents} 

\RequirePackage{xspace}

\DeclareRobustCommand*{\etal}{et al.\xspace}
\DeclareRobustCommand*{\vs}{{\it vs.\ }}
\DeclareRobustCommand*{\aa}{$\accentset{\circ}{\text{A}}$}

\newcommand\T{\rule{0pt}{2.6ex}}

\begin{document}
\preprint{MSUDFT-2014-01}

\title{Structural, elastic and thermal properties of cementite
  (Fe$_3$C) calculated using Modified Embedded Atom Method}

\author{Laalitha~S.~I.~Liyanage}
\affiliation{%
  Department of Physics and Astronomy, 
  Mississippi State University,
  Mississippi State, MS 39762, USA
}
\affiliation{%
  Center for Advanced Vehicular Systems, 
  Mississippi State University, 
  Mississippi State, MS 39762, USA
}

\author{Seong-Gon~Kim}
\email{Author to whom correspondence should be addressed; kimsg@ccs.msstate.edu}
\affiliation{%
  Department of Physics and Astronomy, 
  Mississippi State University,
  Mississippi State, MS 39762, USA
}
\affiliation{%
  Center for Computational Sciences, 
  Mississippi State University, 
  Mississippi State, MS 39762, USA
}

\author{Jeff~Houze}
\author{Sungho~Kim}
\affiliation{%
  Center for Advanced Vehicular Systems, 
  Mississippi State University, 
  Mississippi State, MS 39762, USA
}

\author{Mark~A.~Tschopp}
\affiliation{%
  Army Research Laboratory,
  Aberdeen Proving Ground, MD 21005, USA
}

\author{M.~I.~Baskes}
\affiliation{%
  Department of Mechanical and Aerospace Engineering, 
  University of California, San Diego,
  La Jolla, CA 92093, USA
}
\affiliation{%
  Los Alamos National Laboratory
  Los Alamos, NM 87545, USA
}

\author{M.~F.~Horstemeyer}
\affiliation{%
  Department of Mechanical Engineering, 
  Mississippi State University,
  Mississippi State, MS 39762, USA
}
\affiliation{%
  Center for Advanced Vehicular Systems, 
  Mississippi State University, 
  Mississippi State, MS 39762, USA
}

\date{\today}

\begin{abstract}
  Structural, elastic and thermal properties of cementite (Fe$_3$C)
  were studied using a Modified Embedded Atom Method (MEAM) potential
  for iron-carbon (Fe--C) alloys. Previously developed Fe and C single
  element potentials were used to develop a Fe--C alloy MEAM potential,
  using a statistically-based optimization scheme to reproduce structural
  and elastic properties of cementite, the interstitial energies of
  C in bcc Fe as well as heat of formation of Fe--C alloys in L$_{12}$ 
  and B$_1$ structures. The stability of cementite was investigated by 
  molecular dynamics simulations at high temperatures. The nine single 
  crystal elastic constants for cementite were obtained by computing 
  total energies for strained cells. Polycrystalline elastic moduli
  for cementite were calculated from the single crystal elastic constants of
  cementite. The formation energies of (001), (010), and (100)
  surfaces of cementite were also calculated. The melting temperature
  and the variation of specific heat and volume with respect to
  temperature were investigated by performing a two-phase
  (solid/liquid) molecular dynamics simulation of cementite. The
  predictions of the potential are in good agreement with
  first-principles calculations and experiments.
\end{abstract}

\maketitle

\section{Introduction}
\label{sec:intro}
Steel alloys are the most widely used structural materials due to
their abundance, all-purpose applicability and low cost. The main
carbide in steel alloys is cementite, which forms a
precipitate. Cementite has a direct impact on the mechanical,
structural, and thermal properties of steel. Therefore the ability to
describe and predict properties of cementite at the nanoscale is
essential in the study and design of new steels. Atomistic simulation
methods, such as molecular dynamics or Monte Carlo simulations, offer
an efficient and reliable route to investigate nanoscale mechanics
pertaining to cementite in steel alloys.  Each of these methods
requires accurate interatomic potentials to find the energy of the
system under investigation. However, first-principles
calculations--albeit rigorous and accurate--are incapable of simulating
a large number of atoms required for realistic calculations due to
unreasonable memory and processing-time requirements. Therefore,
semi-empirical potential methods are being explored as a suitable
alternative.

Among the spectrum of semi-empirical formulations, the Modified
Embedded Atom Method (MEAM)\cite{Baskes1992}, originally proposed by
Baskes \etal, has been shown to accurately predict properties of most
crystal structures, such as bcc, fcc, hcp, and even diatomic gases, in
good agreement with experiments or first-principles calculations.
MEAM is extended from the Embedded Atom Method (EAM)\cite{Daw1984} to
include the directionality of bonds. In the original MEAM formalism,
only the first-nearest neighbor (1NN) interactions were
considered.\cite{Baskes1992} Lee and Baskes later extended the
original formalism to include the screened second-nearest neighbor
(2NN) interactions.\cite{Lee2001} Further details of the MEAM
formalism can be found in Ref.~\onlinecite{Baskes1992, Lee2001}.

One of the commonly used 2NN MEAM potentials for the Fe--C system
developed by Byeong-Joo Lee \cite{Lee2006} is designed to predict the
interactions of interstitial C atoms with defects, such as
vacancies. According to Fang \etal,\cite{Fang2012}
Lee's potential predicts that cementite is only stable up to a
temperature of 750~K.\cite{Fang2012} Experimentally, however,
cementite is metastable with a positive heat of formation
\cite{Meschel1997} and only decomposes between 1100 and 1200
K.\cite{Callister2007, Henriksson2009} Among recent interatomic
potentials \cite{Becquart2007,Lau2007,
  Hepburn2008,Ruda2009,Henriksson2009} for the Fe--C system, EAM
potentials by Lau \etal\cite{Lau2007} and Ruda \etal\cite{Ruda2009}
and the short ranged Tersoff-Brenner type analytical bond order
potential (ABOP) by Henriksson \etal\cite{Henriksson2009} all promise
to predict properties of cementite reasonably well. In the potentials
by Lau \etal\cite{Lau2007} and Ruda \etal,\cite{Ruda2009} however, the
single element potential for C does not predict properties of both
graphite and diamond well. This is due to the limited ability of EAM
to describe the bare C-C interaction correctly.\cite{Andrew2010} We
note that a successful interatomic potential for an alloy system
should not only predict the properties of the alloy correctly, but it
should also predict the properties of the individual alloying elements
in their natural crystal structures accurately. The ABOP by Henriksson
\etal\cite{Henriksson2009} accurately predicts properties of cementite
as well as Fe and C; however, ABOPs are not applicable to simulations
involving interfaces and surfaces.\cite{Erhart2006} Furthermore ABOPs
are restricted to 1NN interactions only.\cite{Erhart2006,Albe2002}
Some of the more recent potentials for the Fe--C system are implemented
using in-house developed molecular dynamics codes, which limits the
potentials' usability by a wide scientific community.

In the present work, we developed a 2NN MEAM potential for the Fe--C
alloy system that predicts the structure and properties of
cementite. Our Fe--C alloy potential is based on previously developed
2NN MEAM potentials for Fe \cite{TLee2012} and C \cite{Uddin2010} in
their pure forms. The C MEAM potential predicts both diamond and
graphite as stable structures with almost degenerate energies. Using
the Fe and C single element potentials, we arrived at the best
possible parameterization of the alloy potential of Fe--C for the
purposes specified by the objective function which takes into account
the various properties of Fe--C alloys.

\section{Methods}
\label{sec:methods}

\subsection{MEAM Calculations}
For all atomistic simulations described in the present work, we used
MEAM as implemented in LAMMPS, the classical molecular dynamics
simulation code from Sandia National Laboratories.\cite{plimpton1995,
 plimpton2007} To compare the results of the current potential with
published potentials of Refs.~\onlinecite{Ruda2009} and
~\onlinecite{Henriksson2009} we used the published data. For an
extensive comparison of all properties of cementite with Lee's
potential\cite{Lee2006} we obtained the LAMMPS version of the
potential from the author and conducted our own calculations using
LAMMPS.

\subsection{DFT calculations}
Some of the reference data required for potential construction and
validation are not readily available from experiments. With respect to
the Fe--C system these include the heat of formation of Fe--C in the
$B_1$ structure, the heat of formation of Fe--C in the $L_{12}$
structure and the interstitial energies of C in the bcc Fe lattice at
octahedral and tetrahedral positions. To obtain these properties, we
performed first-principles calculations using Density Functional
Theory (DFT)\cite{Kresse1993, Kresse1996} with Projector Augmented
Wave (PAW) pseudopotentials.\cite{Kresse1999} Electron exchange and
correlation were treated with the Generalized Gradient Approximation
(GGA) as parameterized by Perdew \etal\cite{Perdew1996} Brillouin
zone sampling was performed using the Monkhorst-Pack
scheme,\cite{Monkhorst1976} with a Fermi-level smearing of 0.2~eV
applied using the Methfessel-Paxton method.\cite{Methfessel1989}
Geometric optimizations were carried out using the conjugate gradient
minimization method.\cite{Kresse1993}

\section{Single Element Potentials}
The single element MEAM potential parameters used in the present work
are presented in Table~\ref{tab:pure_pot}. The parameters for Fe are
from the MEAM potential developed by Lee \etal,\cite{TLee2012} and
the parameters for C are from Uddin \etal\cite{Uddin2010}

\begin{table*}[!htbp]
  \caption{\label{tab:pure_pot} Set of the MEAM potential parameters 
    for pure Fe (by Lee \etal\cite{TLee2012}) and
    C (by Uddin \etal\cite{Uddin2010}).
    The bcc and diamond lattices are chosen as the reference structures 
    for Fe and C, respectively.  See Ref.~\onlinecite{Baskes1992, Lee2001} 
    for the meaing of each parameter.}
    \begin{ruledtabular}
      \begin{tabular}{cccccccccccccccccc}
          Element &
          $E_\text{c}$ & $r_e$ & $r_{\text{cut}}$ & 
          $A$ & $\alpha$ & $a_3$ & $\rho_0$ & 
          $\beta^{(0)}$ & $\beta^{(1)}$ & $\beta^{(2)}$ & $\beta^{(3)}$ &
          $t^{(0)}$ & $t^{(1)}$ & $t^{(2)}$ & $t^{(3)}$ &
          $C_{\text{min}}$ & $C_{\text{max}}$ \\
          \hline
          \T
          Fe   &
          4.28 & 2.469 & 4.5   & 
          0.585 & 5.027 & 0.3  & 1.0 &
          3.8  & 2.0   &  0.9  & 0.0 &
          1.0  & $-0.8$  & 12.3  & 2.0 &
          1.9  & 2.8 \\
          C    &
          7.37 & 1.545 & 4.5   & 
          1.49 & 4.38  & 0.0   & 1.0 &
          4.26 & 5.0   &  3.2  & 3.98 &
          1.0  & 7.5   & 1.04  & $-1.01$ &
          0.68 & 2.0 \\
      \end{tabular}
    \end{ruledtabular}
\end{table*}

\subsubsection{Energy \vs volume curves}

\begin{figure}[!htb]
  \includegraphics[scale=.35,keepaspectratio=true]{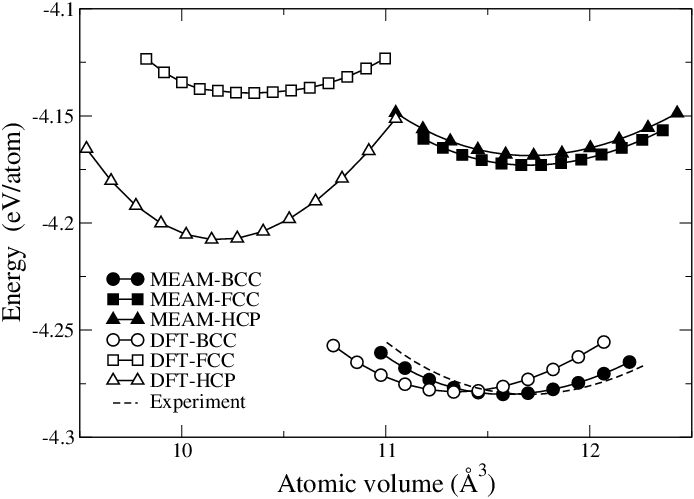}
  \caption{\label{fig:Fe_compare} Energy \vs volume curves for Fe in
    bcc, fcc and hcp crystal structures. The solid curve is
    constructed from experimental values in
    Table~\ref{tab:1elm_props}. For ease of comparison, the DFT curves
    are shifted vertically by a constant amount equal to the difference
    between experimental and DFT cohesive energies of Fe in bcc at
    equilibrium volumes.}
\end{figure}

\begin{figure}[!htb]
  \includegraphics[scale=.35,keepaspectratio=true]{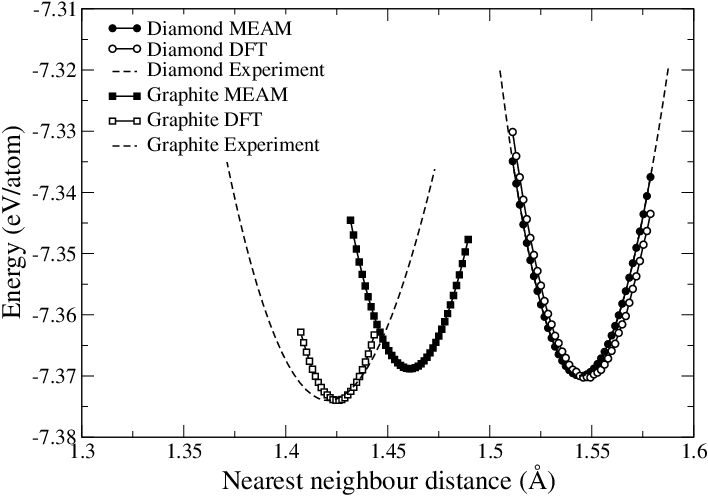}
  \caption{\label{fig:C_EvsnnD} Energy \vs nearest
    neighbor distance curves for C in diamond and graphite.
    The solid curve is constructed from experimental
    values in Table~\ref{tab:1elm_props}. For comparison, the
    DFT curve is shifted vertically to the experimental cohesive
    energy at the equilibrium nearest neighbor distance.}
\end{figure}

\begin{figure}[!htb]
  \includegraphics[scale=.35,keepaspectratio=true]{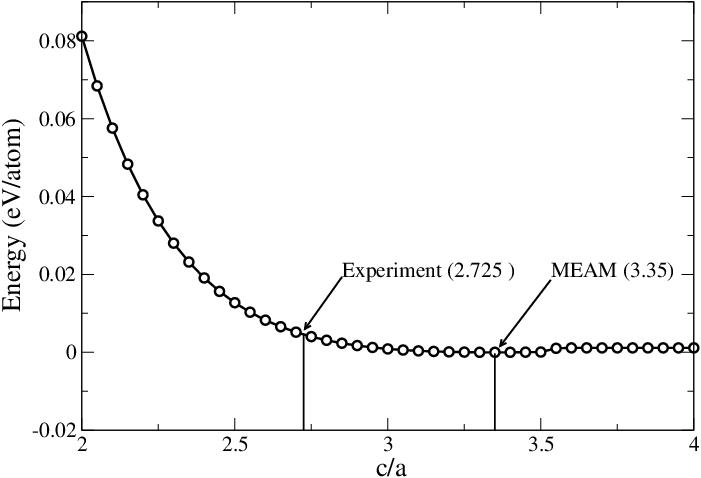}
  \caption{\label{fig:EvsCoa}Cohesive energy of graphite as a function
    of the $c/a$ ratio. Energy at zero is set to the minimum energy
    predicted by MEAM.}
\end{figure}

Energy variation with respect to volume or nearest neighbor distance
is considered an important test of validity for interatomic
potentials. Here we present the energy \vs volume curves generated by
the single element potential for Fe and energy \vs nearest neighbor
distance curves generated by the single element potential for
C. Fig.~\ref{fig:Fe_compare} shows the energy \vs volume curve for
bcc Fe in comparison with curves generated by DFT calculations as well
as using experimental data. It is well known that DFT overestimates
the cohesive energy.\cite{philipsen1996} Therefore, the DFT curve is
shifted vertically by a constant amount to the experimental cohesive
energy at the equilibrium volume to aid the comparison of the
curves. Due to overbinding, the DFT's prediction for the equilibrium
volume is underestimated.\cite{Devey2010} Therefore, the DFT curve
sits to the left of the experimental curve. The experimental curve was
generated through Rose's equation of state \cite{Rose1984}
(Eq.~(\ref{eq:rose})) using the experimental bulk modulus, cohesive
energy, and atomic volume at equilibrium listed in
Table~\ref{tab:1elm_props}. We also tested the stability of Fe in
several different crystal structures including body-centered cubic
(bcc), face-centered cubic (fcc) and hexagonal closed packed (hcp)
structures as shown in Fig.~\ref{fig:Fe_compare}. The Fe MEAM
potential correctly predicts that bcc is the most stable structure, as
observed in experiment and by the first-principles methods.  MEAM
predicts that fcc and hcp Fe are much closer in energy and have a
larger volume than that calculated from DFT.

The single-element MEAM potential for C predicts both diamond and graphite to
be stable structures. Energy \vs nearest neighbor distance curves for
diamond and graphite are shown in Fig.~\ref{fig:C_EvsnnD}. The
experimental curves were constructed from Rose's equation of state
\cite{Rose1984} (Eq.~(\ref{eq:rose})) using the experimental bulk
modulus, cohesive energy, and nearest neighbor distance at
equilibrium, as listed in Table~\ref{tab:1elm_props}. MEAM predictions
for diamond are in good agreement with experiment. MEAM predicts
almost degenerate cohesive energies for graphite and diamond, while
DFT predicts graphite to be $\sim$ 0.1~eV more stable than
diamond. For graphite, DFT predicts a first-nearest neighbor (1NN)
distance in good agreement with experiment, while MEAM predicts a 1NN
distance $\sim$ 3\% greater than the experimental value. The
experimental ratio between lattice parameters $c$ and $a$ in graphite
(hereafter referred as $c/a$ ratio) is 2.725.\cite{Yin1984} MEAM
optimized the $c/a$ ratio of the graphite structure to 3.35. The
disagreement between experimental and MEAM values for $c/a$ ratio is
due to the incorrect prediction of interlayer interaction of graphite,
which is dominated by van der Waals forces that are not described by
MEAM. However, the dependence of cohesive energy on the $c/a$ ratio is
small. Fig.~\ref{fig:EvsCoa} shows the change in energy as $c$ is
varied while keeping $a$ at the MEAM optimized value. According to
Fig.~\ref{fig:EvsCoa}, the difference in cohesive energy of graphite
between the experimental and MEAM $c/a$ ratio is $\sim$ 4~meV/atom. In
constructing the energy \vs nearest neighbor distance curves for
graphite, the inter-planar distance was scaled with the lattice
constant. The experimental ratio was used in the generation of the DFT
curve, while the MEAM curve was constructed with the predicted $c/a$
ratio.

\subsubsection{Single element material properties}
The cohesive energy, equilibrium lattice constants, and bulk moduli
for bcc Fe, graphite, and diamond were determined by fitting Rose's
equation of state \cite{Rose1984}
\begin{align}
\label{eq:rose}
E_{i}^u\left( R \right) &= -E_{i}^0\left(
  1 + a^* + a_3 \frac{{a^{*}}^3}{R/R_{i}^0} \right)e^{-a^{*}}\\
a^{*} &= \alpha_{i} \left( \frac{R}{R_{i}^0} - 1 \right)\\
\label{eq:alpha}
\alpha_{i}^2 &= 9B_i\Omega_i/E_i^0
\end{align}
to the energy \vs nearest neighbor distance/volume curves generated by
MEAM.  $R_i^0$ is the equilibrium nearest neighbor distance, $E_i^0$
is the cohesive energy, $B_i$ is the bulk modulus, $\Omega_i$ is the
equilibrium atomic volume and $a_3$ is the coefficient of the cubic
term.  $a_3$ is set to zero when fitting to energy \vs nearest
neighbor distance/volume curves generated by MEAM.  The single element
material properties compared to experimental values are given in
Table~\ref{tab:1elm_props}.

\begin{table}[!htb]
  \caption{\label{tab:1elm_props} Material properties predicted 
    by the single element MEAM potentials. 
    $E_c$ is the cohesive energy (eV/atom); 
    $a$ and $c$ are the equilibrium lattice constants (\aa);
    $B$ is the bulk modulus (GPa); and 
    $\Omega_0$ is the equilibrium atomic volume (\aa$^3/$atom). 
    Experimental data are given in parentheses. 
    Experimental values for equilibrium atomic volume were 
    calculated from the experimental lattice parameter(s).}
  \begin{ruledtabular}
     \begin{tabular}{cccc}
      Property & bcc Fe & diamond & graphite \\
      \hline
      \T
      $E_c$ &  $-4.28$ ($-4.28$\footnotemark[1]) & 
      $-7.37$ ($-7.37$\footnotemark[3]) & 
      $-7.369$ ($-7.374$\footnotemark[4]) \\
      $B$ & 175 (166-173\footnotemark[2]) & 443 (443\footnotemark[3]) 
      & 176 (286\footnotemark[4]) \\
      $a$ & 2.86 (2.86\footnotemark[2]) & 3.567 (3.567\footnotemark[3])
      & 2.53 (2.461\footnotemark[4]) \\
      $c$ & --- & --- & 8.476 (6.709\footnotemark[4]) \\
      $\Omega_0$ & 11.64 (11.70) & 5.67 (5.67) & 11.75 (8.80)\\
    \end{tabular}
  \end{ruledtabular}
  \footnotetext[1]{Ref.~\onlinecite{Kittel1986} as reported by Lee \etal\cite{TLee2012}}
  \footnotetext[2]{As reported by Lee \etal\cite{TLee2012}}
  \footnotetext[3]{Refs.~\onlinecite{Donohue1982,Mcskimin1972} as reported by Fahy \etal\cite{Fahy1987}}
  \footnotetext[4]{Refs.~\onlinecite{OrsonL1966,murnaghan1944} as reported by Yin \etal \cite{Yin1984}}
\end{table}

\section{Construction of Fe--C Alloy Potential}
\label{sec:meam_alloy}
Table~\ref{tab:alloy_pot} lists the parameters in the 2NN MEAM
potential for Fe--C alloy system optimized by following the general
framework developed by Tschopp \etal\cite{tschopp2011} The framework
consists of two stages. The first stage, called the Global Approach
(GA), is a coarse refinement of the parameter space of the MEAM
potential, which initializes the MEAM potential parameters and
performs a sensitivity analysis for the parameters.  The second stage,
called the Local Approach (LA), evaluates the sensitive parameters
sampling the parameter space with a stratified sampling method and
generates analytical models for design optimization of the potential.

In the GA stage, a coarse refinement of the parameter space is
performed using a partial set of the properties in the objective
function including: the heats of formation of cementite, Fe$_3$C in
$L_{12}$ structure and FeC in $B_1$ structure, and the interstitial
energies of C in the bcc Fe lattice at octahedral and tetrahedral
positions.  The potential parameters were initialized as specified by
the MEAM formulation.\cite{Baskes1992,Lee2001} $\alpha$ defined by
Eq.~(\ref{eq:alpha}) and $r_e$ (equilibrium nearest neighbor distance)
are determined by the reference structure properties.  For the present
case, FeC in $L_{12}$ structure is used as the reference structure and
the values predicted by DFT are used to set $\alpha$ and $r_e$ since
experimental values are not available for this hypothetical structure.
Parameters $\alpha$ and $r_e$ remain unchanged throughout the
optimization process since they are defined by the MEAM formulation.
Next, a sensitivity analysis was performed to evaluate the influence
of each parameter on the properties.  This step helps identify
parameters with the most significant effect on the selected target
properties of the Fe--C system.  By identifying the parameters that
have the most influence on the properties of the Fe--C system we are
able to reduce the number of parameters to be included in the later
stages.  For the present case, the GA stage identified five
parameters---$\Delta$, $a_3$, $\rho_0$(C), $C_{\text{min}}$(Fe,Fe,C),
and $C_{\text{min}}$(C,C,Fe)---to be sufficiently sensitive to be
further explored in the LA stage of the optimization.  Parameters that
are deemed insensitive are fixed at the default values recommended in
the MEAM formulation.

\begin{table}[!htbp]
  \caption{\label{tab:alloy_pot} The optimized parameters in the 2NN MEAM 
    potential for Fe--C alloy system. The triplet ($A$,$B$,$C$) represents 
    the configuration with $C$ atom in between $A$ and $B$ atoms. 
    The $B_1$ lattice is chosen as the reference structure. 
  }
  \begin{ruledtabular}
    \begin{tabular}{cc}
      Parameter & Value \\
      \hline
      \T
      $\Delta$               & 0.002  \\
      $r_e$                  & 1.92   \\
      $r_{\text{cut}}$          & 4.5   \\
      $\alpha$               & 4.75   \\      
      $a_3$                  & 0.125  \\
      $\rho_0$(Fe)           & 1.0    \\
      $\rho_0$(C)            & 5.49   \\
      $C_{\text{max}}$(Fe,Fe,C) & 2.8   \\
      $C_{\text{max}}$(Fe,C,C)  & 2.8   \\      
      $C_{\text{max}}$(Fe,C,Fe) & 2.8   \\
      $C_{\text{max}}$(C,C,Fe)  & 2.8   \\
      $C_{\text{min}}$(Fe,Fe,C) & 0.06  \\
      $C_{\text{min}}$(Fe,C,C)  & 2.0   \\
      $C_{\text{min}}$(Fe,C,Fe) & 2.0   \\
      $C_{\text{min}}$(C,C,Fe)  & 0.5   \\
    \end{tabular}
  \end{ruledtabular}
\end{table}

The LA stage of the potential optimization procedure involves sampling
the bounded potential parameter space, generating analytical models
that represent the nonlinear correlations between the potential
parameters, and using an objective function to converge on the
required parameterization of the potential.  A stratified random
sampling method known as Latin Hypercube Sampling (LHS)
\cite{Mckay2000} was used to sample the potential parameter space with
4000 different potential parameter combinations.  The set of
properties chosen for the Fe--C system are calculated for each
parameter combination.  This is the most computationally-intensive
step of the potential fitting process. Using the data from the
parameter space sampling step, analytical models representing the
relationship between potential parameters and the selected target
properties are generated.  This is done by fitting higher-order
polynomial regression models to the sampled data.  The analytical
models represent a response surface for the sensitive potential
parameters.  At this stage of the optimization an objective function
representing all of the interested properties of the Fe--C system is
introduced.  The objective function is constructed by combining the
weighted differences between the MEAM predicted values and the target
values of the chosen properties.  Target values are set to
experimental values when available or DFT values otherwise.  Then a
constrained nonlinear optimization procedure is used to evaluate the
analytical models by minimizing the objective function.

The properties included in the objective function are the properties
of cementite (equilibrium lattice parameters and volume, heat of
formation, elastic constants, and surface formation energies);
properties of Fe$_3$C in $L_{12}$ structure (heat of formation and
equilibrium volume); properties of FeC in $B_1$ structure (heat of
formation, equilibrium volume and elastic constants) and interstitial
defect energies of C in the bcc Fe lattice at octahedral and
tetrahedral positions.  The weighting factors of the objective
function balance the trade-offs in potential optimization.  The
purpose of the present work is to model the properties of cementite
while reproducing the Fe--C alloy system properties to an acceptable
accuracy.  This is realized by choosing weighting factors in a way
that cementite properties were prioritized first, then the
interstitial defect energies, and then the properties of hypothetical
structures $B_1$ and $L_{12}$.  By varying the weights, the objective
function is changed and the constrained nonlinear optimization
procedure can traverse the response surface represented by the
analytical models to obtain a final set of potential parameters.  For
each set of weighting factors a potential is generated.  By using a
matrix of weighting factors with the required prioritization of the
target properties, we were able to minimize the objective function and
arrive at the set of optimal potential parameters in
Table~\ref{tab:alloy_pot}.  The optimized potential is then validated
by predicting material properties that were not used in the
optimization procedure.  We used the melting temperature of cementite
to validate the potential and its prediction is explained in
Sec.~\ref{sec:cmt_meltTemp}.  Table~\ref{tab:alloy_prop} shows the
material properties predicted by the present MEAM potential compared
with DFT/experimental data and the values from other existing
potentials.

\begin{table*}[!htbp]
  \caption{\label{tab:alloy_prop} Comparison of the present MEAM potential 
    with DFT/experimental data and potentials by Lee\cite{Lee2006}, 
    Ruda \etal \cite{Ruda2009} and Henriksson \etal\cite{Henriksson2009} 
    $\Delta H_{\text{f}}$ is the heat of formation, 
    $\Omega_0$ is the equilibrium volume, 
    $B$ is polycrystalline bulk modulus, 
    $G$ is polycrystalline shear modulus,
    $Y$ is polycrystalline Young's modulus and 
    $\nu$ is polycrystalline Poisson's ratio.
  }
  \begin{ruledtabular}
    \resizebox{\textwidth}{!}{
      \begin{tabular}{lcccccc}
        Properties & DFT/Expt. & MEAM & Lee\cite{Lee2006} & Ruda\cite{Ruda2009} & Henriksson\cite{Henriksson2009}\\
        \hline
        &&&&&\\
        Cementite &&&&&\\
        \hline
        \T
        $\Delta H_{\text{f}}$ (eV/atom)      & 0.01 (0.05\footnotemark[1])& 0.06  & 0.02\footnotemark[9],-0.015\footnotemark[10]  & 0.18 & 0.03 \\
        $\Omega_0$ (\aa$^{3}/\text{atom}$)   & 9.56\footnotemark[2] (9.67\footnotemark[12])& 9.49  & 9.50  & 9.11 & 9.33 \\
        Lattice parameters (\aa)&&&&&&\\
        $a$  & 5.06\footnotemark[2] (5.08\footnotemark[12]) & 5.05 & 5.16  & 5.14 & 5.09 \\
        $b$  & 6.70\footnotemark[2] (6.73\footnotemark[12]) & 6.69 & 6.32  & 6.52 & 6.52 \\
        $c$  & 4.51\footnotemark[2] (4.52\footnotemark[12]) & 4.49 & 4.66  & 4.35 & 4.50 \\
        Elastic constants (GPa) &&&&&&\\
        $C_{11}$                   & 388\footnotemark[3] & 322 &  & 263  & 363\\ 
        $C_{22}$                   & 345\footnotemark[3] & 232 &  & 219  & 406\\
        $C_{33}$                   & 322\footnotemark[3] & 326 &  & 247  & 388\\
        $C_{12}$                   & 156\footnotemark[3] & 137 &  & 176  & 181\\
        $C_{23}$                   & 162\footnotemark[3] & 118 &  & 143  & 130\\
        $C_{13}$                   & 164\footnotemark[3] & 170 &  & 146  & 166\\
        $C_{44}$                   & 15\footnotemark[3]  & 17  &  &  77  &  91\\
        $C_{55}$                   & 134\footnotemark[3] & 103 &  &  95  & 125\\
        $C_{66}$                   & 134\footnotemark[3] & 64  &  & 123  & 134\\    
        Polycrystalline moduli &&&&&&\\
        $B$ (GPa)          & 224 (174$\pm 6$\footnotemark[4]) & 188 & & 183 & 234 \\
        $G$ (GPa)         &  72 (74\footnotemark[5])         & 56 & &  69 & 114 \\
        $Y$ (GPa)         & 194 (177\footnotemark[6], 196\footnotemark[7],200\footnotemark[5]) & 153 & & 184 & 293 \\
        $\nu$        & 0.36 (0.36\footnotemark[5]) & 0.36 & & 0.33 & 0.29 \\
        Surface energies (J/m$^2$) &&&&&&\\
        $E_{(001)}$         & 2.05\footnotemark[8] & 2.05 &  & 1.96 &\\
        $E_{(010)}$         & 2.26\footnotemark[8] & 1.80 &  & 2.00 &\\ 
        $E_{(100)}$         & 2.47\footnotemark[8] & 2.01 &  & 2.34 &\\
        &&&&&\\
        Interstitial Energies (eV)&&&&&&\\
        \hline
        \T
        $E_{\text{Tetrahedral}}$       & 2.14    & 1.76  &  & 2.08 & 1.50\\
        $E_{\text{Octahedral}}$       & 1.25    & 1.55  &  & 1.81 & 1.18\\
        &&&&&\\
        Hypothetical structures &&&&&&\\
        \hline
        \T
        \T
        $\Delta H_{\text{f}}$ $B_1$ (eV/atom)           &   0.53   &  0.002  &  & &\\
        $\Omega_0 B_1$ (\aa$^{3}/\text{atom}$)
                                                        &   7.97   &  7.08   & 8.49 & &\\
        $\Delta H_{\text{f}}$ $L_{12}$ (eV/atom)        &   0.72   &  0.66   &  & &\\
        $\Omega_0 L_{12}$ (\aa$^{3}/\text{atom}$)
                                                        &   10.27   &  10.05   & & &\\
        &&&&&\\
        $B_1$ elastic constants (GPa) &&&&&\\
        $C_{11}$                        & 601   & 566 & 550\footnotemark[11] & & & \\
        $C_{12}$                        & 589   & 213 & 228\footnotemark[11] & & & \\
        $C_{44}$                        &  83   & 145 & 33\footnotemark[11] & & & \\
        \end{tabular}   
        }         
  \end{ruledtabular}
\footnotetext[1]{Meschel \etal \cite{Meschel1997}}
\footnotetext[2]{Shein \etal \cite{Shein2006}}               
\footnotetext[3]{Data from relaxed calculations done by Jiang \etal \cite{Jiang2008}}
\footnotetext[4]{Li \etal \cite{Li2002}}
\footnotetext[5]{Laszlo \etal \cite{Laszlo1959}}
\footnotetext[6]{Mizubayashi \etal \cite{Mizubayashi1999}}
\footnotetext[7]{Umemoto \etal \cite{Umemoto2001}}
\footnotetext[8]{Chiou \etal \cite{Chiou2003}}
\footnotetext[9]{B.-J Lee \cite{Lee2006}}
\footnotetext[10]{Fang \etal \cite{Fang2012}}
\footnotetext[11]{Private communication with B.-J. Lee}
\footnotetext[12]{Wood \etal \cite{Wood2004} as cited by Shein \etal \cite{Shein2006}}
\end{table*}

\section{Structural and Elastic Properties of Cementite}
\begin{figure}[!htbp]
  \includegraphics[scale=.35,keepaspectratio=true]{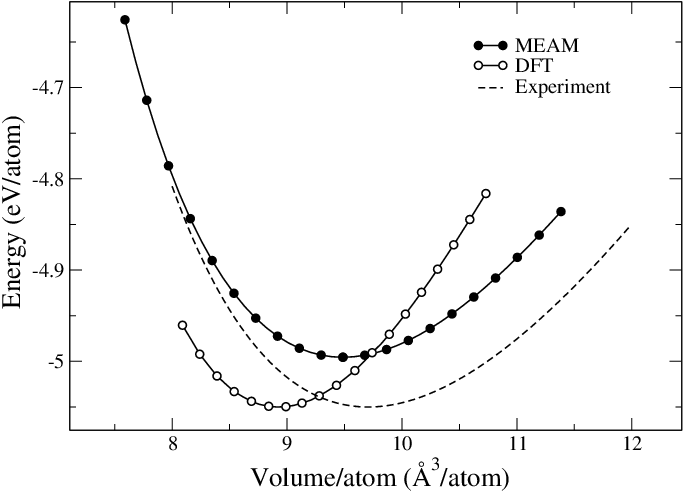}
  \caption{\label{fig:cmt_EvsV} Comparison of energy \vs volume curves
    for cementite. The dashed-line curve is constructed from experimental
    values of the cohesive energy, equilibrium volume and polycrystalline 
    bulk modulus, of cementite.  For comparison, the DFT curve is
    shifted vertically to the experimental cohesive energy at the
    equilibrium volume.}
\end{figure}

Structural properties of cementite including the equilibrium lattice
parameters, the equilibrium volume per atom, and the heat of formation
are presented in Table~\ref{tab:alloy_prop} with comparison to
DFT/experiment, and other interatomic potentials.  Our prediction of
the heat of formation of cementite is in good agreement with DFT and
experimental data.  Lee's and Henriksson's potentials also predict
values in good agreement with DFT and experiment, while Ruda's
potential predicts a much larger value.  Lattice constants of the
present MEAM and literature potentials \cite{Lee2006, Ruda2009,
 Henriksson2009} agree well with experiment, while DFT predicts lower
values.  As a test of validity, the variation of cohesive energy with
volume was calculated.  Fig.~\ref{fig:cmt_EvsV} compares the energy
\vs volume curves for cementite generated by the present MEAM
potential with DFT and experimental curves.  During volume variation of
cementite, the ratios between $a$, $b$ and $c$ lattice parameters were
held constant.  As noted before, DFT overestimates the cohesive energy
and underestimates the equilibrium volume.  Therefore, the DFT curve
sits to the left of the experimental curve, and it is shifted
vertically to the experimental cohesive energy at the equilibrium
volume to aid the comparison.  The experimental curve was generated by
Murnaghan's equation of state\cite{murnaghan1944, murnaghan1967}
\begin{align}
  E\left(V\right) =\ &E(V_0) +
  \frac {B_0 V} {B_0' (B_0' - 1)}\nonumber\\
  &\times
  \left [
    B_0' \left(1 - \frac{V_0}{V}\right)
    + \left(\frac{V_0}{V}\right) ^{B_0'}
    - 1
  \right ].
  \label{eq:murn}
\end{align}
with the experimental bulk modulus $B_0$ \cite{Li2002}, it's
derivative $B_0'$ \cite{Li2002}, volume $V_0$ \cite{Umemoto2001}, and
cohesive energy $E(V_0)$.\cite{Meschel1997} The experimental
single-crystal bulk modulus of cementite has not yet been determined;
therefore, the polycrystalline bulk modulus of cementite was used to
generate the experimental curve.

\subsection{Single-Crystal Elastic Properties}
The elastic moduli of cementite were calculated and compared to DFT
and the interatomic potentials by Ruda \etal,\cite{Ruda2009} and
Henriksson \etal\cite{Henriksson2009} as presented in
Table~\ref{tab:alloy_prop}.  They were calculated using the deformation
matrix presented in Jiang \etal\cite{Jiang2008} In linear elastic
theory, deformation energy is a function of strain.  Distortion
energies ($\Delta E$) calculated for strains ($\delta$) equal to $\pm
0.5$\% were fitted to $\Delta E = k_2 \delta^2 + k_3 \delta^3$.  DFT
calculations were performed for $\delta=\pm 2$\%.\cite{Jiang2008} The
single-crystal elastic constants were obtained using the relationships
for the quadratic coefficient $k_2$ listed in Jiang
\etal\cite{Jiang2008} These results show that the present MEAM
potential for Fe--C alloy predicts cementite to be stable (positive
elastic constants) and their values are reasonably close to those
predicted by DFT.  Specifically, the present MEAM potential reproduces
the low value of $c_{44}$ reported by DFT, which none of the other
interatomic potentials were able to do (MEAM $c_{44}$ of 17 GPa vs DFT
$c_{44}$ 15 GPa).

\subsection{Polycrystalline Elastic Properties}
Theoretical upper and lower bounds for the polycrystalline bulk
modulus ($B$) and shear modulus ($G$) were calculated using the
single-crystal elastic constants according to methods by Reuss and
Voigt.\cite{panda2006, Jiang2008} The polycrystalline $B$ and $G$
were then estimated using Hill's average.\cite{hill1952, Jiang2008}
Young's modulus ($Y$) and Poisson's ratio ($\nu$) were calculated by
using:\cite{Jiang2008}
\begin{eqnarray}
  \label{eq:ymod}
  Y=9BG/(3B+G)\\
  \label{eq:pratio}
  \nu=(3B/2-G)/(3B+G).
\end{eqnarray}

Polycrystalline elastic moduli predicted by the present MEAM potential
are presented in Table~\ref{tab:alloy_prop}, in comparison with DFT,
experiment, and interatomic potentials by Ruda \etal,\cite{Ruda2009}
and Henriksson \etal\cite{Henriksson2009} The elastic constants
predicted by DFT are in good agreement with experiment.  The present
MEAM potential gives the best agreement with experiment among the
three interatomic potentials for $B$ and $\nu$; the present MEAM
predicts the $\nu$ value equal to the experimental value.  Ruda's
potential predicts the best agreement with experiment for $G$ and $Y$.

\subsection{Surface Energies}
Calculations were performed on (001), (010), and (100) surfaces to
determine the surface formation energy.  Table~\ref{tab:alloy_prop}
compares the surface formation energies of the present MEAM to DFT
\cite{Chiou2003} and the interatomic potential by Ruda
\etal\cite{Ruda2009} The atoms near the surfaces are fully relaxed to
allow reconstruction if necessary.  The predicted surface energies
have the same order of magnitude as DFT results.  However, the present
MEAM gives a wrong order of stability among the three surfaces.

\section{Interstitial Energies}
The interstitial point defect formation energy
$E_{\text{f}}^{\text{int}}$ is given by
\begin{equation}
  E_{\text{f}}^{\text{int}} = E_{\text{tot}}[N+A] - E_{\text{tot}}[N] 
  - \epsilon_{\text{A}}
  \label{eq:Ef_int}
\end{equation}
where the total energy of a system with $N$ (Fe or C) atoms is
$E_\text{tot}[N]$ and $E_\text{tot}[N+A]$ is the total energy of a
system with $N$ atoms plus the inserted atom A (Fe or C), and
$\epsilon_{\text{A}}$ is the total energy per atom of type-$A$ in its
most stable bulk structure.  In this case, we considered interstitial
defects of C atoms in a Fe bcc lattice.  Interstitial defect formation
energies of C at the octahedral and tetrahedral positions of the Fe
bcc lattice were calculated.  The results are presented in
Table~\ref{tab:alloy_prop} with comparison to DFT results, and to other
interatomic potentials.  The present MEAM potential predicts the
octahedral defect to be the most stable in agreement with DFT
results.  However, the difference between two defect energies is smaller
compared to that of DFT.

\section{Properties of Hypothetical Crystal Structures}
Heat of formation of Fe--C in $B_1$ crystal structure and $L_{12}$
crystal structure as well as their equilibrium volumes are also
presented in Table~\ref{tab:alloy_prop}. The heat of formation of
$B_1$ is unusually low compared to DFT results. $B_1$ is the reference
structure of the Fe--C alloy potential and its heat of formation is
defined by the $\Delta$ parameter of the potential. The $\Delta$
parameter also has a large effect on the heat of formation of
cementite and thereby to its structural and elastic properties.  Heat
of formation of $B_1$ and $L_{12}$ were used as target properties in
the GA stage of the potential construction process.  However the heat
of formation of these two structures were weighted far less in the
construction of the objective function for obtaining the final
potential parameters as compared to properties of cementite. This
caused the $\Delta$ parameter to have a low value to reproduce overall
cementite properties with greater accuracy. This should not pose a
serious problem since $B_1$ is a hypothetical structure.

\subsubsection{Energy \vs volume curves for $B_1$ and $L_{12}$
  structures}

\begin{figure}[!htb]
  \includegraphics[scale=.35,keepaspectratio=true]{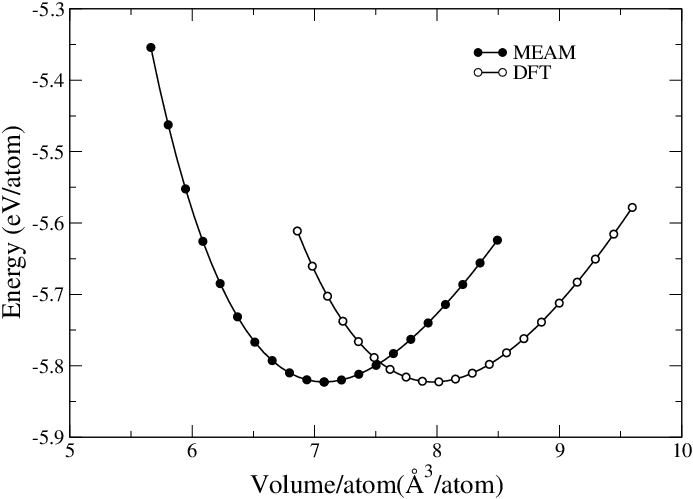}
  \caption{\label{fig:B1_EvsnnD} Comparison of the energy \vs volume
    curves of Fe--C alloy system in the $B_1$ structure. DFT curve is
    shifted vertically to the MEAM-predicted cohesive energy at the
    equilibrium nearest neighbor distance to aid the comparison with
    the MEAM curve.}
\end{figure}

\begin{figure}[!htb]
  \includegraphics[scale=.35,keepaspectratio=true]{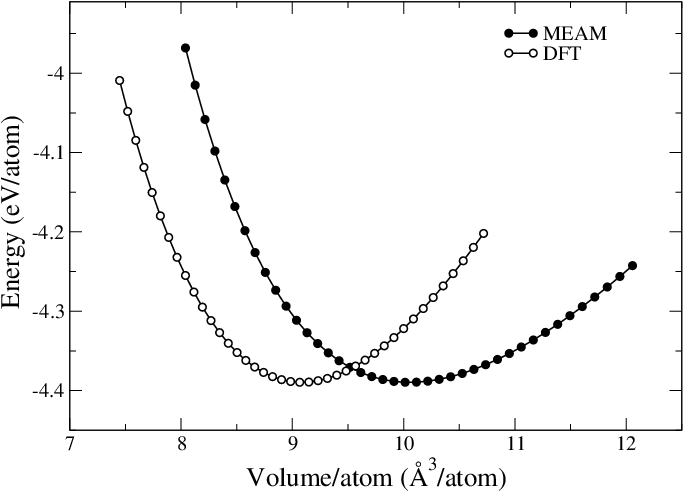}
  \caption{\label{fig:L12_EvsnnD} Comparison of the energy \vs volume
    curves of Fe--C alloy system in the $L_{12}$ structure. DFT curve
    is shifted vertically to the MEAM-predicted cohesive energy at the
    equilibrium nearest neighbor distance aid the comparison with the
    MEAM curve.}
\end{figure}

The cohesive energy of Fe--C in the $B_1$ and $L_{12}$ crystal
structures as a function of the atomic volume is shown in
Figs.~\ref{fig:B1_EvsnnD} and ~\ref{fig:L12_EvsnnD}, respectively.
For the $B_1$ structure, the present MEAM potential predicts an atomic
volume $\sim$ 11\% less, and a bulk modulus $\sim$ 0.3\% less than
DFT.  The MEAM prediction for the $L_{12}$ structure gives an atomic
volume $\sim$ 11\% greater, and a bulk modulus 35\% less than DFT.  As
mentioned earlier, DFT overestimates the cohesive energy.  Therefore,
to aid the comparison in these figures, the DFT curves are shifted
vertically by constant amounts to the MEAM-predicted cohesive energies
at the equilibrium nearest neighbor distances.

\subsubsection{Elastic constants of FeC in the $B_1$ crystal
  structure}

Elastic constants of Fe--C in the $B_1$ crystal structure were
calculated using the Fe--C MEAM potential and are listed in
Table~\ref{tab:alloy_prop} in comparison with DFT calculations and the
interatomic potential by Lee.\cite{Lee2006} They were calculated using
the deformation matrix presented by Jiang \etal\cite{Jiang2008}
Distortion energies ($\Delta E$) calculated for strains ($\delta$)
equal to $\pm 0.1$\% were fitted to $\Delta E = k_2 \delta^2 + k_3
\delta^3$.  The result from the present work for $C_{11}$ compares
reasonably well with the DFT result. $C_{12}$ is predicted at a lower
value than DFT, but it is in the same order of magnitude.  MEAM
prediction of $C_{44}$ is significantly larger than the DFT result.

\section{Thermal properties of cementite}
\label{sec:cmt_meltTemp}

\subsection{Thermal stability of cementite}

The stability of cementite at high temperatures was investigated
through molecular dynamics (MD) simulations in a canonical (NVT)
ensemble from temperatures ranging from 300~K to 1400~K.  At the end
of these MD simulations, cementite retained its crystalline structure,
affirming its stability at high temperatures. The present Fe--C MEAM
potential was also used to predict several thermal properties of
cementite.  In this section, we present calculations for predicting
melting temperature and variation of specific heat and volume of
cementite with respect to temperature.

\subsection{Melting temperature simulation}

Cementite does not have a well-defined melting temperature due to its
metastable nature.\cite{Henriksson2009} Experimentally, cementite
decomposes to ferrite (bcc Fe) and graphite if heated for between 923
K and 973 K for several years.\cite{Callister2007} The Fe--C phase
diagram also has well-defined eutectic point at 1420~K,
\cite{Callister2007,Okamoto1992} where liquid consisting of Fe and C
austenite (fcc Fe) and cementite co-exists in equilibrium.  Above
1420~K the phase diagram for cementite is determined through
mathematical calculations.\cite{Okamoto1992} For the purpose of this
calculation, we considered the melting temperature of cementite to be
the temperature when cementite loses its crystal structure and becomes
a random collection of Fe and C atoms.  The melting temperature
calculation can be done using a single-phase simulation box.  However,
the single phase method generally overestimates the melting
temperature due to the lack of the interface
effects.\cite{belonoshko1994} To avoid this superheating problem and
predict the melting temperature more accurately, we used a two-phase
simulation box that contains both solid and liquid phases.

\begin{figure*}[!htb]
\begin{tabular}{ccc}
\includegraphics[scale=.33,keepaspectratio=true]{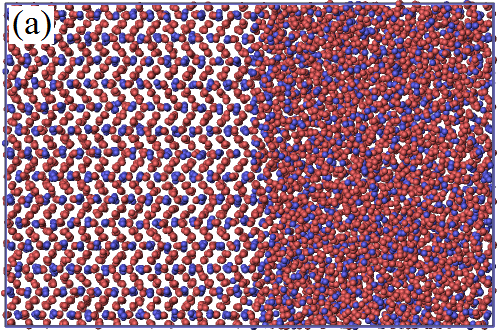} \hspace{5mm}&
\includegraphics[scale=.33,keepaspectratio=true]{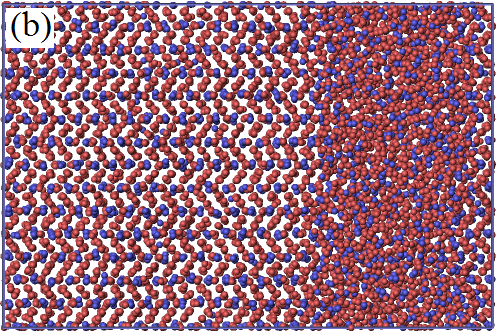} \hspace{5mm}&
\includegraphics[scale=.33,keepaspectratio=true]{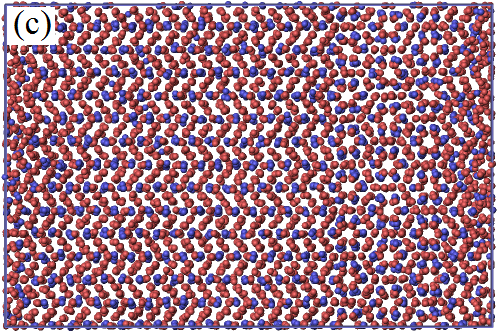} \\
\end{tabular}
\caption{\label{fig:1420K} (Color online) Snapshots of the two-phase
  MD simulation in the NPT ensemble with $T = 1420~K$ and $P = 0$. Red
  spheres are Fe atoms and blue spheres are C atoms. (a) Initial state
  of the simulation box, which contains both liquid and solid phases
  of cementite. (b) Intermediate state of the simulation box at 16~ns,
  as the solid phase propagates to the liquid phase. (c) Final state
  of the simulation box at 32~ns, when the entire system has turned
  into a solid phase. \\}

\begin{tabular}{ccc}
\includegraphics[scale=.33,keepaspectratio=true]{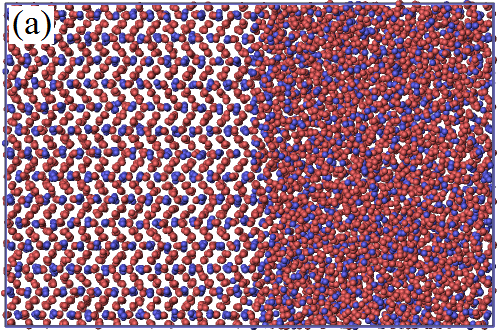} \hspace{5mm}&
\includegraphics[scale=.33,keepaspectratio=true]{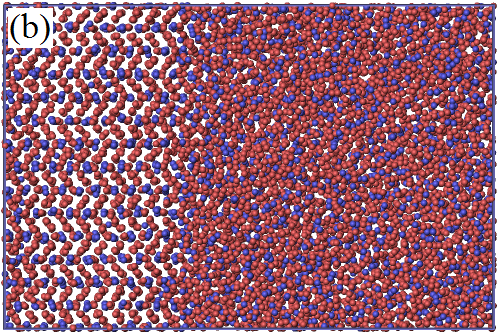}  \hspace{1mm}&
\includegraphics[scale=.33,keepaspectratio=true]{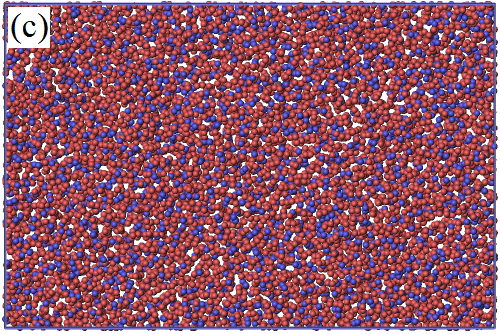} 
\end{tabular}
\caption{\label{fig:1430K} (Color online) Snapshots of the two-phase
  MD simulation in the NPT ensemble with $T = 1430~K$ and $P=0$. Red
  spheres are Fe atoms and blue spheres are C atoms. (a) Initial state
  of the simulation box, which contains both liquid and solid phases
  of cementite. (b) Intermediate state of the simulation box at 20~ns,
  as the liquid phase propagates to the solid phase. (c) Final state
  of the simulation box at 30~ns, when the entire system has turned
  into a liquid phase. }

\end{figure*}

\subsubsection{Preparation of two-phase simulation box}

We performed two-phase simulations (TPS) in the isothermal-isobaric
(NPT) ensemble to determine the melting temperature of cementite.  The
simulation box contained both solid and liquid phases of cementite.
First a supercell containing $14 \times 7 \times 7$ unit cells of
cementite ($10976$ atoms) was heated via MD runs in the NPT ensemble
with $T = 1200$~K and $P = 0$.  Next, one half of the atoms in the
supercell were fixed in their positions and MD runs were carried out
for the other half in the NPT ensemble with a sufficiently high
temperature (such as $T = 4000$~K) and $P = 0$ to create a liquid
phase. The resulting supercell was then subjected to MD runs in the
NPT ensemble with $T = 1500$~K (which is higher than the expected
melting temperature) and $P = 0$, still keeping the same half of the
atoms fixed.  The result of this process was a supercell containing
solid cementite at 1200~K in one half, and liquid cementite at 1500~K
in the other.  This ensures a minimum difference of stress between
atoms in liquid and solid phases of the supercell.  This supercell was
then used in the simulations of solidification and melting of
cementite.

\subsubsection{Two-phase simulation}

The two-phase supercell prepared in the previous section was heated by
MD runs in the NPT ensemble where the temperature $T$ was increased
from 1000~K to 1500~K in 100~K intervals.  Each system ran for 1.6~ns
of simulation time at a time step of 2~fs.  The phase change of the
two-phase simulation box was visually monitored. At 1400~K the solid
phase of the simulation box progressed to occupy the entire box.  In
comparison, at 1500~K the liquid phase of the simulation box
progressed to occupy the entire box.  Next, the initial two-phase
simulation box was heated from 1400~K to 1500~K in 10~K intervals
using NPT MD runs. Each system was equilibrated for at least $5\times
10^6$ time steps totaling to 10~ns.  The final state of the system was
visually inspected.  If the final state appeared to have both liquid
and solid phases, more MD runs were performed until the final state of
the supercell contained only one phase.  Some systems required as much
as 32~ns of MD runs to arrive at a single phase.  The transformation
of the two-phase simulation box to a one-phase simulation box near the
predicted melting temperature is presented in Fig.~\ref{fig:1420K} and
Fig.~\ref{fig:1430K}.  The total energy, volume, and pressure of the
systems were determined through averaging the values of the final
$40\,000$ time steps (80~ps) of each simulation.

\begin{figure*}[!htb]
  \begin{tabular}{ll}
    \includegraphics[scale=.32,keepaspectratio=true]{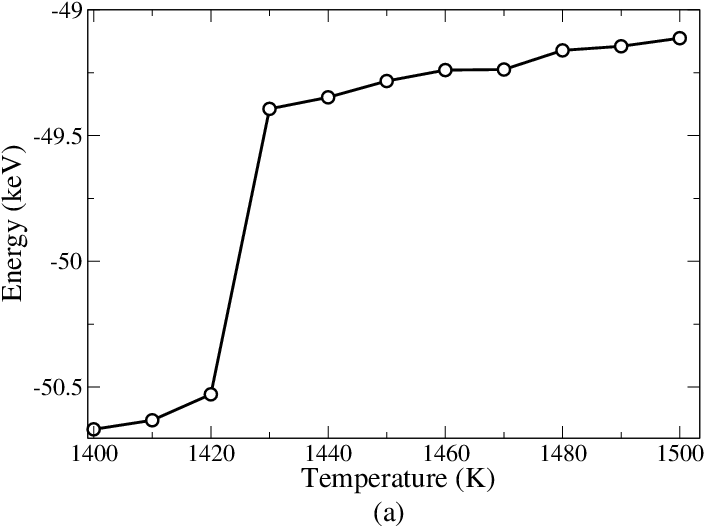} &
    \includegraphics[scale=.32,keepaspectratio=true]{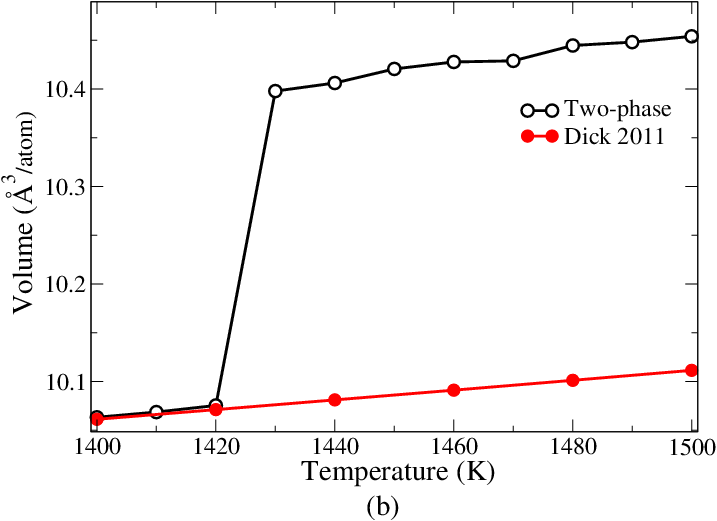} \\
    \includegraphics[scale=.32,keepaspectratio=true]{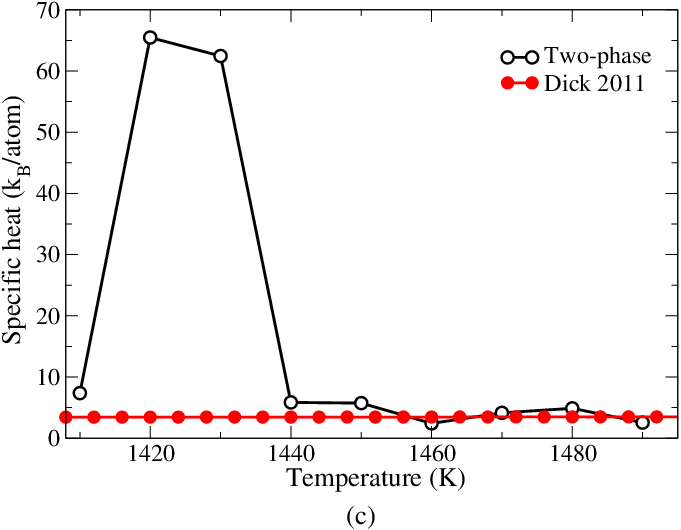} &
    \includegraphics[scale=.32,keepaspectratio=true]{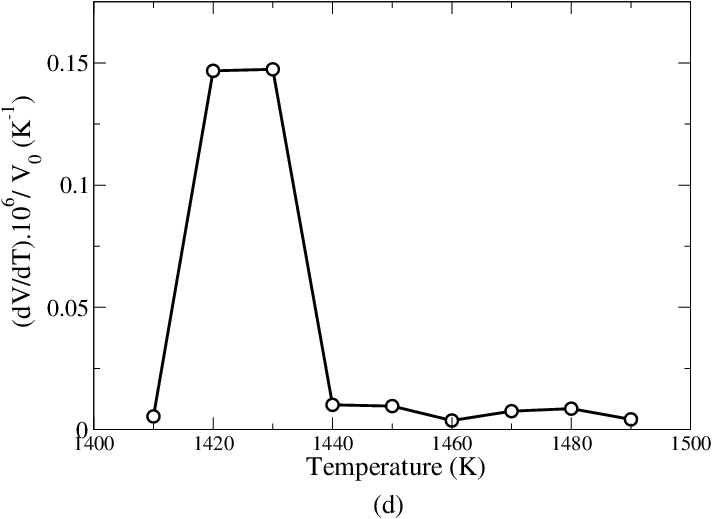} 
  \end{tabular}
  \caption{\label{fig:mt_plots} Variation of properties of the
    two-phase system over the temperature. (a) Total energy of the
    system. (b) (Color online) Volume of the system. Red curve is
    first-principles data by Dick \etal\cite{Dick2011} (c) (Color
    online) Specific heat of the system. Red curve is the
    first-principles data. (d) $dV/dT$ of the system.}
\end{figure*}

In Fig.~\ref{fig:mt_plots} we plot the total energy, volume, specific
heat, and the derivative of volume as functions of temperature.
Experimental data for specific heat and volume are not available for
the 1400-1500~K temperature range. Available experimental data are the
heat capacity of 3.6~$k_B/\text{atom}$ at 1023~K,\cite{naeser1934},
and the experimental volume of 10~\aa$^3/$atom at
1070~K.\cite{Reed1997} Specific heat and volume determined by Dick
\etal from the first-principles calculations\cite{Dick2011} done on
the solid phase of cementite are included for comparison in
Figs.~\ref{fig:mt_plots}(b) and (c).  Since Dick and coworkers used a
single-phase simulation box, their simulation clearly shows
superheating causing the melting temperature to be overestimated.
This can be attributed to the absence of the solid-liquid interface in
single phase simulations. In Fig.~\ref{fig:mt_plots}(c) the specific
heat shows a peak between 1420~K and 1430~K. Therefore we assign
$1425\pm 5$~K as the melting temperature of cementite.  This is a
reasonable prediction compared to experimental eutectic point at
1420~K.\cite{Callister2007,Okamoto1992}

\section{Summary and Conclusion}
In this study, we investigated the properties of cementite using an
interatomic potential developed within the MEAM formalism.  Previously
developed single-element interatomic potentials for Fe and C were used
to develop the Fe--C alloy MEAM potential.  The single-element
potential for C predicts graphite and diamond as stable structures
with nearly degenerate energies.  MEAM potentials for pure elements
predict the heat of formation, bulk moduli, and lattice constants of
Fe and C in their natural crystal structures in good agreement with
experimental data.  The alloy potential for the Fe--C system was
developed primarily to reproduce structural and elastic properties of
cementite.  Secondarily, the interstitial energies of C in bcc Fe, as
well as heats of formation of Fe--C alloys in $B_1$ and $L_{12}$
structures were included with less weighting factors.  The constructed
potential was used to predict structural, elastic, and thermal
properties of cementite. Structural properties tested included the
heat of formation, the equilibrium lattice constants, the equilibrium
volume, and the energy variation with respect to volume.  MEAM
predictions are in good agreement with DFT and experiment.  Nine
single-crystal elastic constants were calculated and used to estimate
polycrystalline bulk modulus, shear modulus, Young's modulus, and
Poisson's ratio of cementite. Surface energies for (001), (010), and
(100) surfaces were also calculated and compared.  The potential was
validated by predicting the thermal stability of cementite, its
melting temperature and the variation of specific heat and volume of
cementite with respect to temperature by two-phase (solid/liquid) MD
simulations.  The present MEAM potential predicted the melting
temperature of cementite to be $1425\pm 5$~K.

\section{Acknowledgments}  

We are grateful to A.B. Belonoshko for his suggestions in conducting
the two-phase melting simulations.  This work was supported in part by
the Department of Energy, grants DE-EE0002323 and DE-FC26-06NT2755.
Computer time allocation has been provided by the High Performance
Computing Collaboratory (HPC$^2$) at Mississippi State University.

\newpage 
\bibliographystyle{apsrev4-1}
\bibliography{FeC,DFT} 

\begin{thebibliography}{53}%
\makeatletter
\providecommand \@ifxundefined [1]{%
 \@ifx{#1\undefined}
}%
\providecommand \@ifnum [1]{%
 \ifnum #1\expandafter \@firstoftwo
 \else \expandafter \@secondoftwo
 \fi
}%
\providecommand \@ifx [1]{%
 \ifx #1\expandafter \@firstoftwo
 \else \expandafter \@secondoftwo
 \fi
}%
\providecommand \natexlab [1]{#1}%
\providecommand \enquote  [1]{``#1''}%
\providecommand \bibnamefont  [1]{#1}%
\providecommand \bibfnamefont [1]{#1}%
\providecommand \citenamefont [1]{#1}%
\providecommand \href@noop [0]{\@secondoftwo}%
\providecommand \href [0]{\begingroup \@sanitize@url \@href}%
\providecommand \@href[1]{\@@startlink{#1}\@@href}%
\providecommand \@@href[1]{\endgroup#1\@@endlink}%
\providecommand \@sanitize@url [0]{\catcode `\\12\catcode `\$12\catcode
  `\&12\catcode `\#12\catcode `\^12\catcode `\_12\catcode `\%12\relax}%
\providecommand \@@startlink[1]{}%
\providecommand \@@endlink[0]{}%
\providecommand \url  [0]{\begingroup\@sanitize@url \@url }%
\providecommand \@url [1]{\endgroup\@href {#1}{\urlprefix }}%
\providecommand \urlprefix  [0]{URL }%
\providecommand \Eprint [0]{\href }%
\providecommand \doibase [0]{http://dx.doi.org/}%
\providecommand \selectlanguage [0]{\@gobble}%
\providecommand \bibinfo  [0]{\@secondoftwo}%
\providecommand \bibfield  [0]{\@secondoftwo}%
\providecommand \translation [1]{[#1]}%
\providecommand \BibitemOpen [0]{}%
\providecommand \bibitemStop [0]{}%
\providecommand \bibitemNoStop [0]{.\EOS\space}%
\providecommand \EOS [0]{\spacefactor3000\relax}%
\providecommand \BibitemShut  [1]{\csname bibitem#1\endcsname}%
\let\auto@bib@innerbib\@empty
\bibitem [{\citenamefont {Baskes}(1992)}]{Baskes1992}%
  \BibitemOpen
  \bibfield  {author} {\bibinfo {author} {\bibfnamefont {M.~I.}\ \bibnamefont
  {Baskes}},\ }\href {\doibase 10.1103/PhysRevB.46.2727} {\bibfield  {journal}
  {\bibinfo  {journal} {Phys. Rev. B}\ }\textbf {\bibinfo {volume} {46}},\
  \bibinfo {pages} {2727} (\bibinfo {year} {1992})}\BibitemShut {NoStop}%
\bibitem [{\citenamefont {Daw}\ and\ \citenamefont {Baskes}(1984)}]{Daw1984}%
  \BibitemOpen
  \bibfield  {author} {\bibinfo {author} {\bibfnamefont {M.~S.}\ \bibnamefont
  {Daw}}\ and\ \bibinfo {author} {\bibfnamefont {M.~I.}\ \bibnamefont
  {Baskes}},\ }\href {\doibase 10.1103/PhysRevB.29.6443} {\bibfield  {journal}
  {\bibinfo  {journal} {Phys. Rev. B}\ }\textbf {\bibinfo {volume} {29}},\
  \bibinfo {pages} {6443} (\bibinfo {year} {1984})}\BibitemShut {NoStop}%
\bibitem [{\citenamefont {Lee}\ \emph {et~al.}(2001)\citenamefont {Lee},
  \citenamefont {Baskes}, \citenamefont {Kim},\ and\ \citenamefont
  {Cho}}]{Lee2001}%
  \BibitemOpen
  \bibfield  {author} {\bibinfo {author} {\bibfnamefont {B.-J.}\ \bibnamefont
  {Lee}}, \bibinfo {author} {\bibfnamefont {M.}~\bibnamefont {Baskes}},
  \bibinfo {author} {\bibfnamefont {H.}~\bibnamefont {Kim}}, \ and\ \bibinfo
  {author} {\bibfnamefont {Y.~K.}\ \bibnamefont {Cho}},\ }\href {\doibase
  10.1103/PhysRevB.64.184102} {\bibfield  {journal} {\bibinfo  {journal} {Phys.
  Rev. B}\ }\textbf {\bibinfo {volume} {64}},\ \bibinfo {pages} {184102}
  (\bibinfo {year} {2001})}\BibitemShut {NoStop}%
\bibitem [{\citenamefont {Lee}(2006)}]{Lee2006}%
  \BibitemOpen
  \bibfield  {author} {\bibinfo {author} {\bibfnamefont {B.}~\bibnamefont
  {Lee}},\ }\href {\doibase 10.1016/j.actamat.2005.09.034} {\bibfield
  {journal} {\bibinfo  {journal} {Acta Materialia}\ }\textbf {\bibinfo {volume}
  {54}},\ \bibinfo {pages} {701} (\bibinfo {year} {2006})}\BibitemShut
  {NoStop}%
\bibitem [{\citenamefont {Fang}\ \emph {et~al.}(2012)\citenamefont {Fang},
  \citenamefont {van Huis}, \citenamefont {Thijsse},\ and\ \citenamefont
  {Zandbergen}}]{Fang2012}%
  \BibitemOpen
  \bibfield  {author} {\bibinfo {author} {\bibfnamefont {C.~M.}\ \bibnamefont
  {Fang}}, \bibinfo {author} {\bibfnamefont {M.~A.}\ \bibnamefont {van Huis}},
  \bibinfo {author} {\bibfnamefont {B.~J.}\ \bibnamefont {Thijsse}}, \ and\
  \bibinfo {author} {\bibfnamefont {H.~W.}\ \bibnamefont {Zandbergen}},\ }\href
  {\doibase 10.1103/PhysRevB.85.054116} {\bibfield  {journal} {\bibinfo
  {journal} {Phys. Rev. B}\ }\textbf {\bibinfo {volume} {85}},\ \bibinfo
  {pages} {054116} (\bibinfo {year} {2012})}\BibitemShut {NoStop}%
\bibitem [{\citenamefont {Meschel}\ and\ \citenamefont
  {Kleppa}(1997)}]{Meschel1997}%
  \BibitemOpen
  \bibfield  {author} {\bibinfo {author} {\bibfnamefont {S.}~\bibnamefont
  {Meschel}}\ and\ \bibinfo {author} {\bibfnamefont {O.}~\bibnamefont
  {Kleppa}},\ }\href {\doibase 10.1016/S0925-8388(97)00023-6} {\bibfield
  {journal} {\bibinfo  {journal} {Journal of Alloys and Compounds}\ }\textbf
  {\bibinfo {volume} {257}},\ \bibinfo {pages} {227 } (\bibinfo {year}
  {1997})}\BibitemShut {NoStop}%
\bibitem [{\citenamefont {Callister}\ and\ \citenamefont
  {Rethwisch}(2007)}]{Callister2007}%
  \BibitemOpen
  \bibfield  {author} {\bibinfo {author} {\bibfnamefont {W.}~\bibnamefont
  {Callister}}\ and\ \bibinfo {author} {\bibfnamefont {D.}~\bibnamefont
  {Rethwisch}},\ }\href@noop {} {\emph {\bibinfo {title} {Materials science and
  engineering: an introduction}}},\ \bibinfo {edition} {7th}\ ed.\ (\bibinfo
  {publisher} {Wiley New York},\ \bibinfo {year} {2007})\ pp.\ \bibinfo {pages}
  {290--293}\BibitemShut {NoStop}%
\bibitem [{\citenamefont {Henriksson}\ and\ \citenamefont
  {Nordlund}(2009)}]{Henriksson2009}%
  \BibitemOpen
  \bibfield  {author} {\bibinfo {author} {\bibfnamefont {K.}~\bibnamefont
  {Henriksson}}\ and\ \bibinfo {author} {\bibfnamefont {K.}~\bibnamefont
  {Nordlund}},\ }\href {\doibase 10.1103/PhysRevB.79.144107} {\bibfield
  {journal} {\bibinfo  {journal} {Physical Review B}\ }\textbf {\bibinfo
  {volume} {79}},\ \bibinfo {pages} {144107} (\bibinfo {year}
  {2009})}\BibitemShut {NoStop}%
\bibitem [{\citenamefont {Becquart}\ \emph {et~al.}(2007)\citenamefont
  {Becquart}, \citenamefont {Raulot}, \citenamefont {Bencteux}, \citenamefont
  {Domain}, \citenamefont {Perez}, \citenamefont {Garruchet},\ and\
  \citenamefont {Nguyen}}]{Becquart2007}%
  \BibitemOpen
  \bibfield  {author} {\bibinfo {author} {\bibfnamefont {C.}~\bibnamefont
  {Becquart}}, \bibinfo {author} {\bibfnamefont {J.}~\bibnamefont {Raulot}},
  \bibinfo {author} {\bibfnamefont {G.}~\bibnamefont {Bencteux}}, \bibinfo
  {author} {\bibfnamefont {C.}~\bibnamefont {Domain}}, \bibinfo {author}
  {\bibfnamefont {M.}~\bibnamefont {Perez}}, \bibinfo {author} {\bibfnamefont
  {S.}~\bibnamefont {Garruchet}}, \ and\ \bibinfo {author} {\bibfnamefont
  {H.}~\bibnamefont {Nguyen}},\ }\href {\doibase
  10.1016/j.commatsci.2006.11.005} {\bibfield  {journal} {\bibinfo  {journal}
  {Computational Materials Science}\ }\textbf {\bibinfo {volume} {40}},\
  \bibinfo {pages} {119} (\bibinfo {year} {2007})}\BibitemShut {NoStop}%
\bibitem [{\citenamefont {Lau}\ \emph {et~al.}(2007)\citenamefont {Lau},
  \citenamefont {F\"orst}, \citenamefont {Lin}, \citenamefont {Gale},
  \citenamefont {Yip},\ and\ \citenamefont {Vliet}}]{Lau2007}%
  \BibitemOpen
  \bibfield  {author} {\bibinfo {author} {\bibfnamefont {T.~T.}\ \bibnamefont
  {Lau}}, \bibinfo {author} {\bibfnamefont {C.~J.}\ \bibnamefont {F\"orst}},
  \bibinfo {author} {\bibfnamefont {X.}~\bibnamefont {Lin}}, \bibinfo {author}
  {\bibfnamefont {J.~D.}\ \bibnamefont {Gale}}, \bibinfo {author}
  {\bibfnamefont {S.}~\bibnamefont {Yip}}, \ and\ \bibinfo {author}
  {\bibfnamefont {K.~J.~V.}\ \bibnamefont {Vliet}},\ }\href {\doibase
  10.1103/PhysRevLett.98.215501} {\bibfield  {journal} {\bibinfo  {journal}
  {Phys. Rev. Lett.}\ }\textbf {\bibinfo {volume} {98}},\ \bibinfo {pages}
  {215501} (\bibinfo {year} {2007})}\BibitemShut {NoStop}%
\bibitem [{\citenamefont {Hepburn}\ and\ \citenamefont
  {Ackland}(2008)}]{Hepburn2008}%
  \BibitemOpen
  \bibfield  {author} {\bibinfo {author} {\bibfnamefont {D.~J.}\ \bibnamefont
  {Hepburn}}\ and\ \bibinfo {author} {\bibfnamefont {G.~J.}\ \bibnamefont
  {Ackland}},\ }\href {\doibase 10.1103/PhysRevB.78.165115} {\bibfield
  {journal} {\bibinfo  {journal} {Phys. Rev. B}\ }\textbf {\bibinfo {volume}
  {78}},\ \bibinfo {pages} {165115} (\bibinfo {year} {2008})}\BibitemShut
  {NoStop}%
\bibitem [{\citenamefont {Ruda}\ \emph {et~al.}(2009)\citenamefont {Ruda},
  \citenamefont {Farkas},\ and\ \citenamefont {Garcia}}]{Ruda2009}%
  \BibitemOpen
  \bibfield  {author} {\bibinfo {author} {\bibfnamefont {M.}~\bibnamefont
  {Ruda}}, \bibinfo {author} {\bibfnamefont {D.}~\bibnamefont {Farkas}}, \ and\
  \bibinfo {author} {\bibfnamefont {G.}~\bibnamefont {Garcia}},\ }\href
  {\doibase 10.1016/j.commatsci.2008.11.020} {\bibfield  {journal} {\bibinfo
  {journal} {Computational Materials Science}\ }\textbf {\bibinfo {volume}
  {45}},\ \bibinfo {pages} {550} (\bibinfo {year} {2009})}\BibitemShut
  {NoStop}%
\bibitem [{\citenamefont {Duff}\ and\ \citenamefont
  {Sluiter}(2010)}]{Andrew2010}%
  \BibitemOpen
  \bibfield  {author} {\bibinfo {author} {\bibfnamefont {A.~I.}\ \bibnamefont
  {Duff}}\ and\ \bibinfo {author} {\bibfnamefont {M.~H.~F.}\ \bibnamefont
  {Sluiter}},\ }\href {http://dx.doi.org/10.2320/matertrans.M2009418}
  {\bibfield  {journal} {\bibinfo  {journal} {MATERIALS TRANSACTIONS}\ }\textbf
  {\bibinfo {volume} {51}},\ \bibinfo {pages} {675} (\bibinfo {year}
  {2010})}\BibitemShut {NoStop}%
\bibitem [{\citenamefont {Erhart}\ \emph {et~al.}(2006)\citenamefont {Erhart},
  \citenamefont {Juslin}, \citenamefont {Goy}, \citenamefont {Nordlund},
  \citenamefont {Müller},\ and\ \citenamefont {Albe}}]{Erhart2006}%
  \BibitemOpen
  \bibfield  {author} {\bibinfo {author} {\bibfnamefont {P.}~\bibnamefont
  {Erhart}}, \bibinfo {author} {\bibfnamefont {N.}~\bibnamefont {Juslin}},
  \bibinfo {author} {\bibfnamefont {O.}~\bibnamefont {Goy}}, \bibinfo {author}
  {\bibfnamefont {K.}~\bibnamefont {Nordlund}}, \bibinfo {author}
  {\bibfnamefont {R.}~\bibnamefont {Müller}}, \ and\ \bibinfo {author}
  {\bibfnamefont {K.}~\bibnamefont {Albe}},\ }\href
  {http://stacks.iop.org/0953-8984/18/i=29/a=003} {\bibfield  {journal}
  {\bibinfo  {journal} {Journal of Physics: Condensed Matter}\ }\textbf
  {\bibinfo {volume} {18}},\ \bibinfo {pages} {6585} (\bibinfo {year}
  {2006})}\BibitemShut {NoStop}%
\bibitem [{\citenamefont {Albe}\ \emph {et~al.}(2002)\citenamefont {Albe},
  \citenamefont {Nordlund},\ and\ \citenamefont {Averback}}]{Albe2002}%
  \BibitemOpen
  \bibfield  {author} {\bibinfo {author} {\bibfnamefont {K.}~\bibnamefont
  {Albe}}, \bibinfo {author} {\bibfnamefont {K.}~\bibnamefont {Nordlund}}, \
  and\ \bibinfo {author} {\bibfnamefont {R.~S.}\ \bibnamefont {Averback}},\
  }\href {\doibase 10.1103/PhysRevB.65.195124} {\bibfield  {journal} {\bibinfo
  {journal} {Phys. Rev. B}\ }\textbf {\bibinfo {volume} {65}},\ \bibinfo
  {pages} {195124} (\bibinfo {year} {2002})}\BibitemShut {NoStop}%
\bibitem [{\citenamefont {Lee}\ \emph {et~al.}(2012)\citenamefont {Lee},
  \citenamefont {Baskes}, \citenamefont {Valone},\ and\ \citenamefont
  {Doll}}]{TLee2012}%
  \BibitemOpen
  \bibfield  {author} {\bibinfo {author} {\bibfnamefont {T.}~\bibnamefont
  {Lee}}, \bibinfo {author} {\bibfnamefont {M.~I.}\ \bibnamefont {Baskes}},
  \bibinfo {author} {\bibfnamefont {S.~M.}\ \bibnamefont {Valone}}, \ and\
  \bibinfo {author} {\bibfnamefont {J.~D.}\ \bibnamefont {Doll}},\ }\href
  {http://stacks.iop.org/0953-8984/24/i=22/a=225404} {\bibfield  {journal}
  {\bibinfo  {journal} {Journal of Physics: Condensed Matter}\ }\textbf
  {\bibinfo {volume} {24}},\ \bibinfo {pages} {225404} (\bibinfo {year}
  {2012})}\BibitemShut {NoStop}%
\bibitem [{\citenamefont {Uddin}\ \emph {et~al.}(2010)\citenamefont {Uddin},
  \citenamefont {Baskes}, \citenamefont {Srinivasan}, \citenamefont {Cundari},\
  and\ \citenamefont {Wilson}}]{Uddin2010}%
  \BibitemOpen
  \bibfield  {author} {\bibinfo {author} {\bibfnamefont {J.}~\bibnamefont
  {Uddin}}, \bibinfo {author} {\bibfnamefont {M.~I.}\ \bibnamefont {Baskes}},
  \bibinfo {author} {\bibfnamefont {S.~G.}\ \bibnamefont {Srinivasan}},
  \bibinfo {author} {\bibfnamefont {T.~R.}\ \bibnamefont {Cundari}}, \ and\
  \bibinfo {author} {\bibfnamefont {A.~K.}\ \bibnamefont {Wilson}},\ }\href
  {\doibase 10.1103/PhysRevB.81.104103} {\bibfield  {journal} {\bibinfo
  {journal} {Phys. Rev. B}\ }\textbf {\bibinfo {volume} {81}},\ \bibinfo
  {pages} {104103} (\bibinfo {year} {2010})}\BibitemShut {NoStop}%
\bibitem [{\citenamefont {Plimpton}(1995)}]{plimpton1995}%
  \BibitemOpen
  \bibfield  {author} {\bibinfo {author} {\bibfnamefont {S.}~\bibnamefont
  {Plimpton}},\ }\href {\doibase 10.1006/jcph.1995.1039} {\bibfield  {journal}
  {\bibinfo  {journal} {Journal of Computational Physics}\ }\textbf {\bibinfo
  {volume} {117}},\ \bibinfo {pages} {1 } (\bibinfo {year} {1995})}\BibitemShut
  {NoStop}%
\bibitem [{\citenamefont {Plimpton}\ \emph {et~al.}()\citenamefont {Plimpton}
  \emph {et~al.}}]{plimpton2007}%
  \BibitemOpen
  \bibfield  {author} {\bibinfo {author} {\bibfnamefont {S.}~\bibnamefont
  {Plimpton}} \emph {et~al.},\ }\href {http://lammps.sandia.gov/} {\enquote
  {\bibinfo {title} {Lammps-large-scale atomic/molecular massively parallel
  simulator},}\ }\bibinfo {note} {April 2011 version}\BibitemShut {NoStop}%
\bibitem [{\citenamefont {Kresse}\ and\ \citenamefont
  {Hafner}(1993)}]{Kresse1993}%
  \BibitemOpen
  \bibfield  {author} {\bibinfo {author} {\bibfnamefont {G.}~\bibnamefont
  {Kresse}}\ and\ \bibinfo {author} {\bibfnamefont {J.}~\bibnamefont
  {Hafner}},\ }\href {\doibase 10.1103/PhysRevB.47.558} {\bibfield  {journal}
  {\bibinfo  {journal} {Phys. Rev. B}\ }\textbf {\bibinfo {volume} {47}},\
  \bibinfo {pages} {558} (\bibinfo {year} {1993})}\BibitemShut {NoStop}%
\bibitem [{\citenamefont {Kresse}\ and\ \citenamefont
  {Furthm\"uller}(1996)}]{Kresse1996}%
  \BibitemOpen
  \bibfield  {author} {\bibinfo {author} {\bibfnamefont {G.}~\bibnamefont
  {Kresse}}\ and\ \bibinfo {author} {\bibfnamefont {J.}~\bibnamefont
  {Furthm\"uller}},\ }\href {\doibase 10.1103/PhysRevB.54.11169} {\bibfield
  {journal} {\bibinfo  {journal} {Phys. Rev. B}\ }\textbf {\bibinfo {volume}
  {54}},\ \bibinfo {pages} {11169} (\bibinfo {year} {1996})}\BibitemShut
  {NoStop}%
\bibitem [{\citenamefont {Kresse}\ and\ \citenamefont
  {Joubert}(1999)}]{Kresse1999}%
  \BibitemOpen
  \bibfield  {author} {\bibinfo {author} {\bibfnamefont {G.}~\bibnamefont
  {Kresse}}\ and\ \bibinfo {author} {\bibfnamefont {D.}~\bibnamefont
  {Joubert}},\ }\href {\doibase 10.1103/PhysRevB.59.1758} {\bibfield  {journal}
  {\bibinfo  {journal} {Phys. Rev. B}\ }\textbf {\bibinfo {volume} {59}},\
  \bibinfo {pages} {1758} (\bibinfo {year} {1999})}\BibitemShut {NoStop}%
\bibitem [{\citenamefont {Perdew}\ \emph {et~al.}(1996)\citenamefont {Perdew},
  \citenamefont {Burke},\ and\ \citenamefont {Ernzerhof}}]{Perdew1996}%
  \BibitemOpen
  \bibfield  {author} {\bibinfo {author} {\bibfnamefont {J.~P.}\ \bibnamefont
  {Perdew}}, \bibinfo {author} {\bibfnamefont {K.}~\bibnamefont {Burke}}, \
  and\ \bibinfo {author} {\bibfnamefont {M.}~\bibnamefont {Ernzerhof}},\ }\href
  {\doibase 10.1103/PhysRevLett.77.3865} {\bibfield  {journal} {\bibinfo
  {journal} {Phys. Rev. Lett.}\ }\textbf {\bibinfo {volume} {77}},\ \bibinfo
  {pages} {3865} (\bibinfo {year} {1996})}\BibitemShut {NoStop}%
\bibitem [{\citenamefont {Monkhorst}\ and\ \citenamefont
  {Pack}(1976)}]{Monkhorst1976}%
  \BibitemOpen
  \bibfield  {author} {\bibinfo {author} {\bibfnamefont {H.}~\bibnamefont
  {Monkhorst}}\ and\ \bibinfo {author} {\bibfnamefont {J.}~\bibnamefont
  {Pack}},\ }\href {\doibase 10.1103/PhysRevB.13.5188} {\bibfield  {journal}
  {\bibinfo  {journal} {Phys. Rev. B}\ }\textbf {\bibinfo {volume} {13}},\
  \bibinfo {pages} {5188} (\bibinfo {year} {1976})}\BibitemShut {NoStop}%
\bibitem [{\citenamefont {Methfessel}\ and\ \citenamefont
  {Paxton}(1989)}]{Methfessel1989}%
  \BibitemOpen
  \bibfield  {author} {\bibinfo {author} {\bibfnamefont {M.}~\bibnamefont
  {Methfessel}}\ and\ \bibinfo {author} {\bibfnamefont {A.~T.}\ \bibnamefont
  {Paxton}},\ }\href {\doibase 10.1103/PhysRevB.40.3616} {\bibfield  {journal}
  {\bibinfo  {journal} {Phys. Rev. B}\ }\textbf {\bibinfo {volume} {40}},\
  \bibinfo {pages} {3616} (\bibinfo {year} {1989})}\BibitemShut {NoStop}%
\bibitem [{\citenamefont {Philipsen}\ and\ \citenamefont
  {Baerends}(1996)}]{philipsen1996}%
  \BibitemOpen
  \bibfield  {author} {\bibinfo {author} {\bibfnamefont {P.~H.~T.}\
  \bibnamefont {Philipsen}}\ and\ \bibinfo {author} {\bibfnamefont {E.~J.}\
  \bibnamefont {Baerends}},\ }\href {\doibase 10.1103/PhysRevB.54.5326}
  {\bibfield  {journal} {\bibinfo  {journal} {Phys. Rev. B}\ }\textbf {\bibinfo
  {volume} {54}},\ \bibinfo {pages} {5326} (\bibinfo {year}
  {1996})}\BibitemShut {NoStop}%
\bibitem [{\citenamefont {Devey}\ and\ \citenamefont
  {de~Leeuw}(2010)}]{Devey2010}%
  \BibitemOpen
  \bibfield  {author} {\bibinfo {author} {\bibfnamefont {A.}~\bibnamefont
  {Devey}}\ and\ \bibinfo {author} {\bibfnamefont {N.~H.}\ \bibnamefont
  {de~Leeuw}},\ }\href {\doibase 10.1103/PhysRevB.82.235112} {\bibfield
  {journal} {\bibinfo  {journal} {Phys. Rev. B}\ }\textbf {\bibinfo {volume}
  {82}},\ \bibinfo {pages} {235112} (\bibinfo {year} {2010})}\BibitemShut
  {NoStop}%
\bibitem [{\citenamefont {Rose}\ \emph {et~al.}(1984)\citenamefont {Rose},
  \citenamefont {Smith}, \citenamefont {Guinea},\ and\ \citenamefont
  {Ferrante}}]{Rose1984}%
  \BibitemOpen
  \bibfield  {author} {\bibinfo {author} {\bibfnamefont {J.~H.}\ \bibnamefont
  {Rose}}, \bibinfo {author} {\bibfnamefont {J.~R.}\ \bibnamefont {Smith}},
  \bibinfo {author} {\bibfnamefont {F.}~\bibnamefont {Guinea}}, \ and\ \bibinfo
  {author} {\bibfnamefont {J.}~\bibnamefont {Ferrante}},\ }\href {\doibase
  10.1103/PhysRevB.29.2963} {\bibfield  {journal} {\bibinfo  {journal} {Phys.
  Rev. B}\ }\textbf {\bibinfo {volume} {29}},\ \bibinfo {pages} {2963}
  (\bibinfo {year} {1984})}\BibitemShut {NoStop}%
\bibitem [{\citenamefont {Yin}\ and\ \citenamefont {Cohen}(1984)}]{Yin1984}%
  \BibitemOpen
  \bibfield  {author} {\bibinfo {author} {\bibfnamefont {M.~T.}\ \bibnamefont
  {Yin}}\ and\ \bibinfo {author} {\bibfnamefont {M.~L.}\ \bibnamefont
  {Cohen}},\ }\href {\doibase 10.1103/PhysRevB.29.6996} {\bibfield  {journal}
  {\bibinfo  {journal} {Phys. Rev. B}\ }\textbf {\bibinfo {volume} {29}},\
  \bibinfo {pages} {6996} (\bibinfo {year} {1984})}\BibitemShut {NoStop}%
\bibitem [{\citenamefont {Kittel}(1986)}]{Kittel1986}%
  \BibitemOpen
  \bibfield  {author} {\bibinfo {author} {\bibfnamefont {C.}~\bibnamefont
  {Kittel}},\ }\href@noop {} {\emph {\bibinfo {title} {{Introduction to Solid
  State Physics}}}},\ \bibinfo {edition} {6th}\ ed.\ (\bibinfo  {publisher}
  {John Wiley \& Sons, Inc.},\ \bibinfo {address} {New York},\ \bibinfo {year}
  {1986})\BibitemShut {NoStop}%
\bibitem [{\citenamefont {Donohue}(1982)}]{Donohue1982}%
  \BibitemOpen
  \bibfield  {author} {\bibinfo {author} {\bibfnamefont {J.}~\bibnamefont
  {Donohue}},\ }\href@noop {} {\emph {\bibinfo {title} {The structures of the
  elements}}}\ (\bibinfo  {publisher} {R.E. Krieger Pub. Co.},\ \bibinfo
  {address} {Malabar, Fla.},\ \bibinfo {year} {1982})\ p.\ \bibinfo {pages}
  {256}\BibitemShut {NoStop}%
\bibitem [{\citenamefont {McSkimin}\ \emph {et~al.}(1972)\citenamefont
  {McSkimin}, \citenamefont {P.~Andreatch},\ and\ \citenamefont
  {Glynn}}]{Mcskimin1972}%
  \BibitemOpen
  \bibfield  {author} {\bibinfo {author} {\bibfnamefont {H.~J.}\ \bibnamefont
  {McSkimin}}, \bibinfo {author} {\bibfnamefont {J.}~\bibnamefont
  {P.~Andreatch}}, \ and\ \bibinfo {author} {\bibfnamefont {P.}~\bibnamefont
  {Glynn}},\ }\href {\doibase 10.1063/1.1661318} {\bibfield  {journal}
  {\bibinfo  {journal} {Journal of Applied Physics}\ }\textbf {\bibinfo
  {volume} {43}},\ \bibinfo {pages} {985} (\bibinfo {year} {1972})}\BibitemShut
  {NoStop}%
\bibitem [{\citenamefont {Fahy}\ and\ \citenamefont {Louie}(1987)}]{Fahy1987}%
  \BibitemOpen
  \bibfield  {author} {\bibinfo {author} {\bibfnamefont {S.}~\bibnamefont
  {Fahy}}\ and\ \bibinfo {author} {\bibfnamefont {S.~G.}\ \bibnamefont
  {Louie}},\ }\href {\doibase 10.1103/PhysRevB.36.3373} {\bibfield  {journal}
  {\bibinfo  {journal} {Phys. Rev. B}\ }\textbf {\bibinfo {volume} {36}},\
  \bibinfo {pages} {3373} (\bibinfo {year} {1987})}\BibitemShut {NoStop}%
\bibitem [{\citenamefont {Orson}\ and\ \citenamefont
  {Anderson}(1966)}]{OrsonL1966}%
  \BibitemOpen
  \bibfield  {author} {\bibinfo {author} {\bibfnamefont {L.}~\bibnamefont
  {Orson}}\ and\ \bibinfo {author} {\bibnamefont {Anderson}},\ }\href {\doibase
  10.1016/0022-3697(66)90199-5} {\bibfield  {journal} {\bibinfo  {journal}
  {Journal of Physics and Chemistry of Solids}\ }\textbf {\bibinfo {volume}
  {27}},\ \bibinfo {pages} {547 } (\bibinfo {year} {1966})}\BibitemShut
  {NoStop}%
\bibitem [{\citenamefont {{Murnaghan}}(1944)}]{murnaghan1944}%
  \BibitemOpen
  \bibfield  {author} {\bibinfo {author} {\bibfnamefont {F.~D.}\ \bibnamefont
  {{Murnaghan}}},\ }\href {\doibase 10.1073/pnas.30.9.244} {\bibfield
  {journal} {\bibinfo  {journal} {Proceedings of the National Academy of
  Science}\ }\textbf {\bibinfo {volume} {30}},\ \bibinfo {pages} {244}
  (\bibinfo {year} {1944})}\BibitemShut {NoStop}%
\bibitem [{\citenamefont {Tschopp}\ \emph {et~al.}(2012)\citenamefont
  {Tschopp}, \citenamefont {Solanki}, \citenamefont {Baskes}, \citenamefont
  {Gao}, \citenamefont {Sun},\ and\ \citenamefont {Horstemeyer}}]{tschopp2011}%
  \BibitemOpen
  \bibfield  {author} {\bibinfo {author} {\bibfnamefont {M.}~\bibnamefont
  {Tschopp}}, \bibinfo {author} {\bibfnamefont {K.}~\bibnamefont {Solanki}},
  \bibinfo {author} {\bibfnamefont {M.}~\bibnamefont {Baskes}}, \bibinfo
  {author} {\bibfnamefont {F.}~\bibnamefont {Gao}}, \bibinfo {author}
  {\bibfnamefont {X.}~\bibnamefont {Sun}}, \ and\ \bibinfo {author}
  {\bibfnamefont {M.}~\bibnamefont {Horstemeyer}},\ }\href {\doibase
  10.1016/j.jnucmat.2011.08.003} {\bibfield  {journal} {\bibinfo  {journal}
  {Journal of Nuclear Materials}\ }\textbf {\bibinfo {volume} {425}},\ \bibinfo
  {pages} {22 } (\bibinfo {year} {2012})}\BibitemShut {NoStop}%
\bibitem [{\citenamefont {McKay}\ \emph {et~al.}(2000)\citenamefont {McKay},
  \citenamefont {Beckman},\ and\ \citenamefont {Conover}}]{Mckay2000}%
  \BibitemOpen
  \bibfield  {author} {\bibinfo {author} {\bibfnamefont {M.}~\bibnamefont
  {McKay}}, \bibinfo {author} {\bibfnamefont {R.}~\bibnamefont {Beckman}}, \
  and\ \bibinfo {author} {\bibfnamefont {W.}~\bibnamefont {Conover}},\ }\href
  {http://www.jstor.org/stable/1271432} {\bibfield  {journal} {\bibinfo
  {journal} {Technometrics}\ }\textbf {\bibinfo {volume} {42}},\ \bibinfo
  {pages} {55} (\bibinfo {year} {2000})}\BibitemShut {NoStop}%
\bibitem [{\citenamefont {Shein}\ \emph {et~al.}(2006)\citenamefont {Shein},
  \citenamefont {Medvedeva},\ and\ \citenamefont {Ivanovskii}}]{Shein2006}%
  \BibitemOpen
  \bibfield  {author} {\bibinfo {author} {\bibfnamefont {I.}~\bibnamefont
  {Shein}}, \bibinfo {author} {\bibfnamefont {N.}~\bibnamefont {Medvedeva}}, \
  and\ \bibinfo {author} {\bibfnamefont {A.}~\bibnamefont {Ivanovskii}},\
  }\href {\doibase http://dx.doi.org/10.1016/j.physb.2005.10.093} {\bibfield
  {journal} {\bibinfo  {journal} {Physica B: Condensed Matter}\ }\textbf
  {\bibinfo {volume} {371}},\ \bibinfo {pages} {126 } (\bibinfo {year}
  {2006})}\BibitemShut {NoStop}%
\bibitem [{\citenamefont {Jiang}\ \emph {et~al.}(2008)\citenamefont {Jiang},
  \citenamefont {Srinivasan}, \citenamefont {Caro},\ and\ \citenamefont
  {Maloy}}]{Jiang2008}%
  \BibitemOpen
  \bibfield  {author} {\bibinfo {author} {\bibfnamefont {C.}~\bibnamefont
  {Jiang}}, \bibinfo {author} {\bibfnamefont {S.~G.}\ \bibnamefont
  {Srinivasan}}, \bibinfo {author} {\bibfnamefont {A.}~\bibnamefont {Caro}}, \
  and\ \bibinfo {author} {\bibfnamefont {S.~A.}\ \bibnamefont {Maloy}},\ }\href
  {\doibase 10.1063/1.2884529} {\bibfield  {journal} {\bibinfo  {journal}
  {Journal of Applied Physics}\ }\textbf {\bibinfo {volume} {103}},\ \bibinfo
  {pages} {043502} (\bibinfo {year} {2008})}\BibitemShut {NoStop}%
\bibitem [{\citenamefont {Li}\ \emph {et~al.}(2002)\citenamefont {Li},
  \citenamefont {Mao}, \citenamefont {Fei}, \citenamefont {Gregoryanz},
  \citenamefont {Eremets},\ and\ \citenamefont {Zha}}]{Li2002}%
  \BibitemOpen
  \bibfield  {author} {\bibinfo {author} {\bibfnamefont {J.}~\bibnamefont
  {Li}}, \bibinfo {author} {\bibfnamefont {H.~K.}\ \bibnamefont {Mao}},
  \bibinfo {author} {\bibfnamefont {Y.}~\bibnamefont {Fei}}, \bibinfo {author}
  {\bibfnamefont {E.}~\bibnamefont {Gregoryanz}}, \bibinfo {author}
  {\bibfnamefont {M.}~\bibnamefont {Eremets}}, \ and\ \bibinfo {author}
  {\bibfnamefont {C.~S.}\ \bibnamefont {Zha}},\ }\href
  {http://dx.doi.org/10.1007/s00269-001-0224-4} {\bibfield  {journal} {\bibinfo
   {journal} {Physics and Chemistry of Minerals}\ }\textbf {\bibinfo {volume}
  {29}},\ \bibinfo {pages} {166} (\bibinfo {year} {2002})},\ \bibinfo {note}
  {10.1007/s00269-001-0224-4}\BibitemShut {NoStop}%
\bibitem [{\citenamefont {Laszlo}\ and\ \citenamefont
  {Nolle}(1959)}]{Laszlo1959}%
  \BibitemOpen
  \bibfield  {author} {\bibinfo {author} {\bibfnamefont {F.}~\bibnamefont
  {Laszlo}}\ and\ \bibinfo {author} {\bibfnamefont {H.}~\bibnamefont {Nolle}},\
  }\href {\doibase 10.1016/0022-5096(59)90006-7} {\bibfield  {journal}
  {\bibinfo  {journal} {Journal of the Mechanics and Physics of Solids}\
  }\textbf {\bibinfo {volume} {7}},\ \bibinfo {pages} {193 } (\bibinfo {year}
  {1959})}\BibitemShut {NoStop}%
\bibitem [{\citenamefont {Mizubayashi}\ \emph {et~al.}(1999)\citenamefont
  {Mizubayashi}, \citenamefont {Li}, \citenamefont {Yumoto},\ and\
  \citenamefont {Shimotomai}}]{Mizubayashi1999}%
  \BibitemOpen
  \bibfield  {author} {\bibinfo {author} {\bibfnamefont {H.}~\bibnamefont
  {Mizubayashi}}, \bibinfo {author} {\bibfnamefont {S.}~\bibnamefont {Li}},
  \bibinfo {author} {\bibfnamefont {H.}~\bibnamefont {Yumoto}}, \ and\ \bibinfo
  {author} {\bibfnamefont {M.}~\bibnamefont {Shimotomai}},\ }\href {\doibase
  10.1016/S1359-6462(99)00003-2} {\bibfield  {journal} {\bibinfo  {journal}
  {Scripta Materialia}\ }\textbf {\bibinfo {volume} {40}},\ \bibinfo {pages}
  {773 } (\bibinfo {year} {1999})}\BibitemShut {NoStop}%
\bibitem [{\citenamefont {Umemoto}\ \emph {et~al.}(2001)\citenamefont
  {Umemoto}, \citenamefont {Liu}, \citenamefont {Masuyama},\ and\ \citenamefont
  {Tsuchiya}}]{Umemoto2001}%
  \BibitemOpen
  \bibfield  {author} {\bibinfo {author} {\bibfnamefont {M.}~\bibnamefont
  {Umemoto}}, \bibinfo {author} {\bibfnamefont {Z.}~\bibnamefont {Liu}},
  \bibinfo {author} {\bibfnamefont {K.}~\bibnamefont {Masuyama}}, \ and\
  \bibinfo {author} {\bibfnamefont {K.}~\bibnamefont {Tsuchiya}},\ }\href
  {\doibase 10.1016/S1359-6462(01)01016-8} {\bibfield  {journal} {\bibinfo
  {journal} {Scripta Materialia}\ }\textbf {\bibinfo {volume} {45}},\ \bibinfo
  {pages} {391 } (\bibinfo {year} {2001})}\BibitemShut {NoStop}%
\bibitem [{\citenamefont {Chiou}(2003)}]{Chiou2003}%
  \BibitemOpen
  \bibfield  {author} {\bibinfo {author} {\bibfnamefont {W.}~\bibnamefont
  {Chiou}},\ }\href {\doibase 10.1016/S0039-6028(03)00352-2} {\bibfield
  {journal} {\bibinfo  {journal} {Surface Science}\ }\textbf {\bibinfo {volume}
  {530}},\ \bibinfo {pages} {87} (\bibinfo {year} {2003})}\BibitemShut
  {NoStop}%
\bibitem [{\citenamefont {Wood}\ \emph {et~al.}(2004)\citenamefont {Wood},
  \citenamefont {Vo{\v{c}}adlo}, \citenamefont {Knight}, \citenamefont
  {Dobson}, \citenamefont {Marshall}, \citenamefont {Price},\ and\
  \citenamefont {Brodholt}}]{Wood2004}%
  \BibitemOpen
  \bibfield  {author} {\bibinfo {author} {\bibfnamefont {I.~G.}\ \bibnamefont
  {Wood}}, \bibinfo {author} {\bibfnamefont {L.}~\bibnamefont {Vo{\v{c}}adlo}},
  \bibinfo {author} {\bibfnamefont {K.~S.}\ \bibnamefont {Knight}}, \bibinfo
  {author} {\bibfnamefont {D.~P.}\ \bibnamefont {Dobson}}, \bibinfo {author}
  {\bibfnamefont {W.~G.}\ \bibnamefont {Marshall}}, \bibinfo {author}
  {\bibfnamefont {G.~D.}\ \bibnamefont {Price}}, \ and\ \bibinfo {author}
  {\bibfnamefont {J.}~\bibnamefont {Brodholt}},\ }\href {\doibase
  10.1107/S0021889803024695} {\bibfield  {journal} {\bibinfo  {journal}
  {Journal of Applied Crystallography}\ }\textbf {\bibinfo {volume} {37}},\
  \bibinfo {pages} {82} (\bibinfo {year} {2004})}\BibitemShut {NoStop}%
\bibitem [{\citenamefont {Murnaghan}(1967)}]{murnaghan1967}%
  \BibitemOpen
  \bibfield  {author} {\bibinfo {author} {\bibfnamefont {F.}~\bibnamefont
  {Murnaghan}},\ }\href@noop {} {\emph {\bibinfo {title} {Finite deformation of
  an elastic solid}}}\ (\bibinfo  {publisher} {Dover New York},\ \bibinfo
  {year} {1967})\BibitemShut {NoStop}%
\bibitem [{\citenamefont {Panda}\ and\ \citenamefont
  {Chandran}(2006)}]{panda2006}%
  \BibitemOpen
  \bibfield  {author} {\bibinfo {author} {\bibfnamefont {K.}~\bibnamefont
  {Panda}}\ and\ \bibinfo {author} {\bibfnamefont {K.~R.}\ \bibnamefont
  {Chandran}},\ }\href {\doibase 10.1016/j.actamat.2005.12.003} {\bibfield
  {journal} {\bibinfo  {journal} {Acta Materialia}\ }\textbf {\bibinfo {volume}
  {54}},\ \bibinfo {pages} {1641 } (\bibinfo {year} {2006})}\BibitemShut
  {NoStop}%
\bibitem [{\citenamefont {Hill}(1952)}]{hill1952}%
  \BibitemOpen
  \bibfield  {author} {\bibinfo {author} {\bibfnamefont {R.}~\bibnamefont
  {Hill}},\ }\href {http://stacks.iop.org/0370-1298/65/i=5/a=307} {\bibfield
  {journal} {\bibinfo  {journal} {Proceedings of the Physical Society. Section
  A}\ }\textbf {\bibinfo {volume} {65}},\ \bibinfo {pages} {349} (\bibinfo
  {year} {1952})}\BibitemShut {NoStop}%
\bibitem [{\citenamefont {Okamoto}(1992)}]{Okamoto1992}%
  \BibitemOpen
  \bibfield  {author} {\bibinfo {author} {\bibfnamefont {H.}~\bibnamefont
  {Okamoto}},\ }\href {http://dx.doi.org/10.1007/BF02665767} {\bibfield
  {journal} {\bibinfo  {journal} {Journal of Phase Equilibria}\ }\textbf
  {\bibinfo {volume} {13}},\ \bibinfo {pages} {543} (\bibinfo {year} {1992})},\
  \bibinfo {note} {10.1007/BF02665767}\BibitemShut {NoStop}%
\bibitem [{\citenamefont {Belonoshko}(1994)}]{belonoshko1994}%
  \BibitemOpen
  \bibfield  {author} {\bibinfo {author} {\bibfnamefont {A.~B.}\ \bibnamefont
  {Belonoshko}},\ }\href {\doibase 10.1016/0016-7037(94)90265-8} {\bibfield
  {journal} {\bibinfo  {journal} {Geochimica et Cosmochimica Acta}\ }\textbf
  {\bibinfo {volume} {58}},\ \bibinfo {pages} {4039 } (\bibinfo {year}
  {1994})}\BibitemShut {NoStop}%
\bibitem [{\citenamefont {Naeser}(1934)}]{naeser1934}%
  \BibitemOpen
  \bibfield  {author} {\bibinfo {author} {\bibfnamefont {G.}~\bibnamefont
  {Naeser}},\ }\href@noop {} {\bibfield  {journal} {\bibinfo  {journal} {Mitt.
  Kais.-Wilh.-Inst. Eisenforschg}\ }\textbf {\bibinfo {volume} {16}},\ \bibinfo
  {pages} {207} (\bibinfo {year} {1934})},\ \bibinfo {note} {as reported in
  Dick \etal \cite{Dick2011}}\BibitemShut {NoStop}%
\bibitem [{\citenamefont {Reed}\ and\ \citenamefont {Root}(1997)}]{Reed1997}%
  \BibitemOpen
  \bibfield  {author} {\bibinfo {author} {\bibfnamefont {R.~C.}\ \bibnamefont
  {Reed}}\ and\ \bibinfo {author} {\bibfnamefont {J.~H.}\ \bibnamefont
  {Root}},\ }\href {\doibase 10.1016/S1359-6462(97)00438-7} {\bibfield
  {journal} {\bibinfo  {journal} {Scripta Materialia}\ }\textbf {\bibinfo
  {volume} {38}},\ \bibinfo {pages} {95 } (\bibinfo {year} {1997})},\ \bibinfo
  {note} {as reported by Dick \etal \cite{Dick2011}}\BibitemShut {NoStop}%
\bibitem [{\citenamefont {Dick}\ \emph {et~al.}(2011)\citenamefont {Dick},
  \citenamefont {K\"ormann}, \citenamefont {Hickel},\ and\ \citenamefont
  {Neugebauer}}]{Dick2011}%
  \BibitemOpen
  \bibfield  {author} {\bibinfo {author} {\bibfnamefont {A.}~\bibnamefont
  {Dick}}, \bibinfo {author} {\bibfnamefont {F.}~\bibnamefont {K\"ormann}},
  \bibinfo {author} {\bibfnamefont {T.}~\bibnamefont {Hickel}}, \ and\ \bibinfo
  {author} {\bibfnamefont {J.}~\bibnamefont {Neugebauer}},\ }\href {\doibase
  10.1103/PhysRevB.84.125101} {\bibfield  {journal} {\bibinfo  {journal} {Phys.
  Rev. B}\ }\textbf {\bibinfo {volume} {84}},\ \bibinfo {pages} {125101}
  (\bibinfo {year} {2011})}\BibitemShut {NoStop}%
\end{thebibliography}%
\end{document}